\documentclass[a4paper,11pt]{article}
\usepackage{jheppub}
\usepackage{booktabs}
\usepackage{kotex}
\usepackage[dvipsnames]{xcolor}
\usepackage{tikz}
\usepackage{caption}
\usepackage{mathtools}
\usepackage{makecell}
\usepackage{booktabs}
\usepackage{multirow}
\usetikzlibrary{decorations.markings,decorations.pathmorphing,positioning}

\def\a{\alpha}
\def\b{\beta}
\def\g{\gamma}
\def\d{\delta}
\def\e{\epsilon}
\def\m{\mu}
\def\n{\nu}
\def\r{\rho}
\def\S{\Sigma}

\def\CC{\mathcal{C}}
\def\CI{\mathcal{I}}
\def\CD{\mathcal{D}}
\def\CF{\mathcal{F}}
\def\CH{\mathcal{H}}
\def\CN{\mathcal{N}}
\def\CT{\mathcal{T}}
\def\CW{\mathcal{W}}
\def\CL{\mathcal{L}}
\def\CO{\mathcal{O}}

\def\pd{\partial}
\def\WRT{{\rm WRT}}
\def\HF{\mathbf{HF}}
\def\TFTk{{\rm{TFT}}[\vec{k}]}
\def\lg{\langle}
\def\rg{\rangle}
\newcommand{\ket}[1]{\left\vert #1\right\rangle}

\arxivnumber{2511.16380}

\title{\boldmath QFT Realization of Non-Unitary $\mathfrak{sl}(2,\mathbb{C})$ WRT Invariants and Their Galois Conjugations}

\author{Kibok Jeong, Soochang Lee}
\affiliation{Department of Physics and Astronomy $\&$ Center for Theoretical Physics,\\
Seoul National University, 1 Gwanak-ro, Seoul 08826, Korea}

\emailAdd{boki0322@snu.ac.kr}
\emailAdd{physicsmp1217@snu.ac.kr}

\abstract{We propose a field theoretic realization of the non-unitary $\mathfrak{sl}(2,\mathbb{C})$ Witten-Reshetikhin-Turaev Topological Quantum Field Theory(WRT TQFT). The WRT TQFT at the principal root of unity is unitary. It is known to be realized by ${\rm SU}(2)$ Chern-Simons theory. However, the WRT TQFT at a non-principal root of unity is non-unitary. Its field theoretic realization has remained unclear. We propose that such a non-unitary TQFT arises from the topological twist of the 3-dimensional $\mathcal{N}=4$ rank-0 theory constructed by joining multiple $T[{\rm SU}(2)]$ theories. We construct its modular matrices and identify them with those of the WRT TQFT, establishing a concrete relation between the parameters, up to a decoupled unitary TQFT.}

\begin{document}
\maketitle
\flushbottom

\section{Introduction} \label{Sec: Intro}
Seeking a deeper understanding of Topological Quantum Field Theory(TQFT) is worthwhile. It is physically significant because unitary TQFTs naturally appear as an effective description for various gapped IR physics. The ${\rm U}(1)$ Chern-Simons theory for the quantum Hall effect is the representative example. With the anticipation of discovering suitable systems to apply, researchers imagine various TQFTs and study their properties\cite{Wen:2015qwa,Witten:1988hf,Witten:1991we,Atiyah:1988tx,Collier:2023fwi,Dimofte:2011gm,Gang:2021hrd,kim2025afx,Gang:2024wxz,go2025xy,Baek:2025uev}.

Focusing on the 3-dimensional Euclidean spacetimes, TQFTs possess the bulk-boundary correspondence which relates the Modular Tensor Category(MTC) data associated with the 2-dimensional Conformal Field Theory(CFT) living on its boundary with various bulk observables\cite{Turaev:1994nc,Witten:1988hf,Moore:1988uz,Gang:2025ykf,Baek:2025uev}. It allows us to construct the bulk TQFT by establishing the associated modular data. However, such a construction doesn't determine a field theoretic description (QFT in figure~\ref{fig:main problem} below) for the bulk TQFT. We seek to identify it through observation-based conjecture.

\begin{figure}[h]
\centering
\includegraphics[width=.45\textwidth]{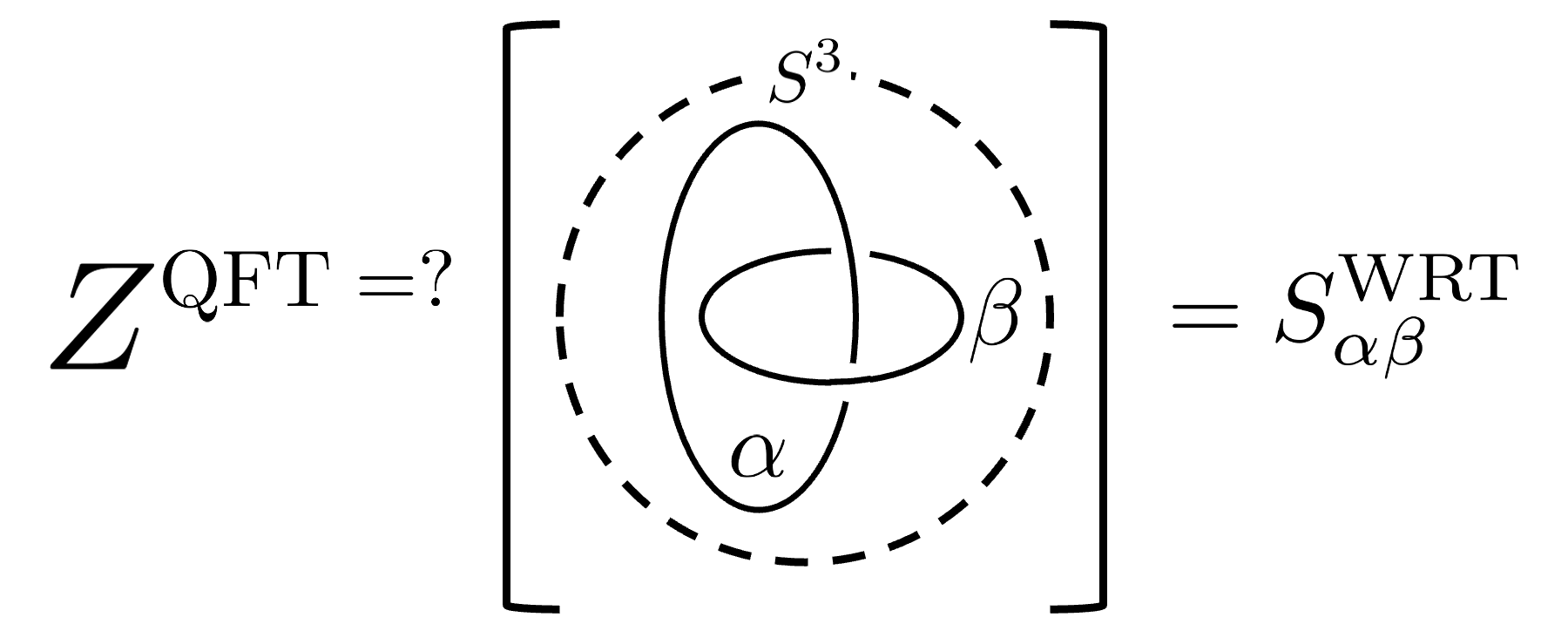}
\captionof{figure}{The problem we are going to deal with in this paper.}
\label{fig:main problem}
\end{figure}

Motivated by the recent surge of interest in non-unitary field theories\cite{Gannon:2003kg,Gang:2021hrd,ElGanainy:2018pcd,Ghatak:2019xvi,Yao:2018aaq,Xiao:2020pil}, we focus on the \(\mathfrak{sl}(2,\mathbb{C})\) Witten-Reshetikhin-Turaev (WRT) TQFT, which arises as a Galois conjugate of the \(\textrm{SU}(2)_k\) Chern-Simons theory. The latter is obtained by applying the Reshetikhin-Turaev construction to the quantum group \(U_x\mathfrak{sl}(2,\mathbb{C})\) at the principal root of unity \(x=e^{\frac{2\pi i}{k+2}}\)(\(m=1\)), and is unitary\cite{Witten:1988hf}. Since the modular data and their defining pentagon and hexagon relations are polynomial with rational coefficients in \(x^{1/4}\), one obtains a Galois-conjugated MTC by rotating \(x\to x^m=e^{\frac{2\pi im}{k+2}}\). This rotation can be realized at two levels (See appendix~\ref{Ap: WRT review}): at the categorical level the modular data lie in \(\mathbb{Q}(\zeta_{4(k+2)})\), so admissible conjugations require \(\gcd(m,4(k+2))=1\), while at the algebraic level the conjugation acts on the root of unity entering the quantum group before the construction, where the weaker condition \(\gcd(m,k+2)=1\) suffices. In particular, even \(m\)-which forces \(k\) odd and gives the \(\textrm{WRT}_k^m\) theories relevant to this work-is reachable only at the algebraic level. Such theories with \(m\geq 2\) are non-unitary\cite{Reshetikhin:1990tk,Reshetikhin:1991tc,Turaev:1994nc}, yet their field theoretic realizations remained unclear. 

Recent studies provide us with a hint. 3-dimensional field theories which possess the $\mathcal{N}=4$ supersymmetry can be non-trivial interacting SCFTs at the IR even though they have empty Coulomb and Higgs branches\cite{Gang:2018huc}. Efforts have been made to classify these so-called \textit{rank-0} theories, leading to the discovery\cite{Gang:2021hrd}
\begin{align*}
    [\text{3d }\mathcal{N}=4\text{ rank-0 SCFT}]\xrightarrow{\text{Top'l twist}}[\text{Non-unitary TQFT}]\;.
\end{align*}

Extending this discovery, properties of the $D(\vec{k})$ theory labeled by $\vec{k}\in(\mathbb{Z^*})^{n>1}$ were studied\cite{Gang:2024wxz,Baek:2025uev}. The $D(\vec{k})$ theory is $\CN=4$ rank-0 SCFT constructed by joining $n-1$ $T[{\rm SU}(2)]$ theories\cite{Terashima:2011qi}, which live on the S-duality domain wall of the 4-dimensional $\mathcal{N}=4$ $\mathrm{SU}(2)$ supersymmetric Yang-Mills theory\cite{Gaiotto:2008sd,Hosomichi:2010vh}. They predicted the following factorization through an anomaly analysis associated with $\mathbb{Z}^{\otimes n}_2$ 1-form symmetry originating from the center subgroups of gauged ${\rm{SU}(2)}^{\otimes n}$ in the $D(\vec{k})$ theory\cite{Hsin:2018vcg}
\begin{align*}
    D(\vec{k})|_{\rm Twist}=\CD(p,q)|_{\rm Twist}\otimes\TFTk\;.
\end{align*}
The first part $\CD(p,q)$ depends on the two coprime integers $p$ and $q$ constructed from the negative continued fraction of $\vec{k}$\footnote{\(\frac{q}{p}=\frac{1}{k_1 - \frac{1}{k_2 - \frac{1}{k_3-\cdots \frac{1}{k_n}}}}\)}. The second part $\TFTk$ is the decoupled (unitary) TQFT of the $\mathrm{U}(1)$ Chern-Simons theory type, whose Chern-Simons level matrix depends on $\vec{k}$\cite{Gang:2024wxz}. They also checked it for some $n=2$ cases.

By applying the intuition gained from the ${\rm U}(1)$ fractional quantum Hall system\cite{Witten:2003ya,Burgess:2001}, it is natural to expect from the definition of the $D(\vec{k})$ theory that it provides a field-theoretic realization of the $\mathfrak{sl}(2,\mathbb{C})$ WRT TQFT. Indeed, the results of \cite{Gang:2024wxz} already show that the modular data of the WRT TQFT are realized as the modular data of the twisted $\CD(p,q)$ theory. Our motivation is explained in detail in section~\ref{Subsec: D(p,q) theory}. In this paper, we  consider the A-twisted $D(\vec{k})$ theory consisting of an arbitrary number of $T[\mathrm{SU}(2)]$ segments. One of the important properties of the $D(\vec{k})$ theory is that the contributions from adjoint chirals in the theory diverge at the A-twist point. This is not an obstruction. It allows us to linearize the Bethe equations, ultimately leading to an analytic solution.

We show that the set of Handle-gluing and Fibering operators, $\mathbf{HF}$ data, is always factorized in the way predicted in \cite{Gang:2024wxz,Hsin:2018vcg}
\begin{align*}
    \mathbf{HF}^{D(\vec{k})}=\mathbf{HF}^{\mathcal{D}(p,q)}\otimes \mathbf{HF}^{\TFTk}
\end{align*}
and identify the first part $\mathbf{HF}^{\mathcal{D}(p,q
)}$ with the $\mathbf{HF}$ data of the WRT TQFT determined by $p$ and $q$. Going further, we investigate the possibility that the tensor product of Wilson lines at the UV becomes a simple object in the IR. When $p$ is odd, we find some candidates for the maximally independent set of simple objects and use them to construct the full modular matrices. The resulting modular matrices are the tensor product of two modular matrices
\begin{align*}
    S,T^{D(\vec{k})}=S,T^{\mathrm{\mathcal{D}}(p,q)}\otimes  S,T^{\mathrm{TFT}[\vec{k}]}
\end{align*}
which strongly supports the prediction of~\cite{Gang:2024wxz,Hsin:2018vcg}. We identify the first part $S,T^{\mathcal{D}(p,q)}$ with modular matrices of the WRT TQFT determined by $p$ and $q$. When $p$ is even, we find a candidate for a subset of simple objects that allows us to construct the $\mathcal{D}(p,q)|_A$ part of modular matrices. We identify the resulting matrices with modular matrices of the WRT TQFT determined by $p$ and $q$. We also argue, using Weyl invariance, that the simple objects we missed cannot be just tensor products of Wilson lines at the UV.

The rest of this paper is organized as follows. In section~\ref{Sec: Field theory}, we provide the motivation to consider the twisted $D(\vec{k})$ theory as a field theoretic realization of the WRT TQFT and present the details of the field theory. We then perform a Bethe/Gauge analysis to extract the data of the non-unitary TQFT $D(\vec{k})|_A$. The definition of the $\HF$ data, which plays a crucial role in this paper, is also defined in this section. See \eqref{eq: def of HF data}. In section~\ref{Sec: WRT data extraction}, we identify the $\CD(p,q)|_A$ with the WRT TQFT determined by $p$ and $q$ through the $\HF$ data analysis and the modular matrix construction using simple objects. In section~\ref{Sec: Discussion}, we discuss some remaining issues and suggest future directions. Appendix~\ref{Ap: WRT review} contains an essential review of the $\mathfrak{sl}(2,\mathbb{C})$ WRT TQFT. Appendix~\ref{Ap: Methods} summarizes the essential background used in the field theory analysis. Computational details, including linearization of Bethe equations, extraction of modular data associated with the $D(\vec{k})|_A$ theory, factorization of $\HF$ data, and determination of full modular matrices, are presented in appendix~\ref{Ap: BG detail} and~\ref{Ap: Full modular matrices}. In appendix~\ref{Ap: Simple obejcts}, we introduce candidates for simple objects of the $D(\vec{k})|_A$ theory we find. Properties of the quantum dilogarithm function we used are summarized in appendix~\ref{Ap: QDL}.

\section{The field theory} \label{Sec: Field theory}
\subsection{$D(\vec{k})$ theory} \label{Subsec: D(k) theory}
\begin{figure}[h]
\centering
\begin{minipage}[b]{0.45\textwidth}
\centering
\includegraphics[width=.75\textwidth]{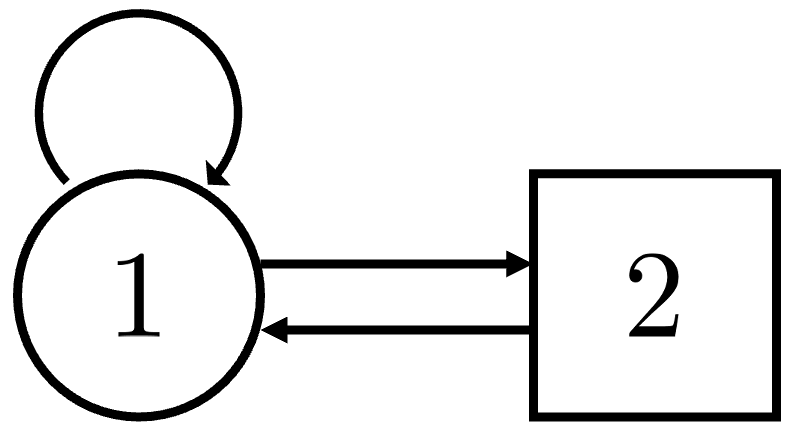}
\captionof{figure}{Quiver diagram associated with the $T[{\rm SU}(2)]$ theory.}
\label{fig: tsu2quiver}
\end{minipage}
\hspace{0.3cm}
\begin{minipage}[b]{0.45\textwidth}
\centering
\begin{tabular}{@{}ccccc@{}}
\toprule
Matters        & $\mathrm{U}(1)_G$          & ${\rm SU}(2)_F$                    & $\mathrm{U}(1)_A$      & $\mathrm{U}(1)_R$      \\ \midrule
$X_1$            & $0$ & $\mathbf{1}$                  & $-1$          & $1$           \\
$\Phi_1$       & $1$               & $\bar{\mathbf{2}}=\mathbf{2}$                  & $\frac{1}{2}$ & $\frac{1}{2}$ \\
$\bar{\Phi}_1$ & $-1$              & $\mathbf{2}$ & $\frac{1}{2}$ & $\frac{1}{2}$ \\ \bottomrule
\end{tabular}
\captionof{table}{Matters in the $T[{\rm SU}(2)]$ theory.}
\label{tab: tsu2content}
\end{minipage}
\end{figure}

Let's consider the $T[{\rm SU}(2)]$ theory, which has matter contents described by a quiver diagram as shown in figure~\ref{fig: tsu2quiver}\cite{Gaiotto:2008sd,Hosomichi:2010vh}. See also table~\ref{tab: tsu2content}. Charges for $\mathrm{U}(1)_A$ and $\mathrm{U}(1)_R$ are restricted by the superpotential ${\bar\Phi}_1 X_1 \Phi_1$.
Additionally, we can introduce the background vector multiplet for the \textit{topological ${\rm U}(1)$} associated with ${\rm U}(1)_G$. This symmetry is known to be enhanced to ${\rm SU}(2)^\vee={\rm SO}(3)$ at the IR\cite{Gaiotto:2008sd,Terashima:2011qi}. However, we can just regard it as ${\rm SU}(2)$ when calculations do not depend on a global structure of Lie group\cite{Terashima:2011qi}. Let's denote the original flavor ${\rm SU}(2)$ as ${\rm SU}(2)_L$ and the ${\rm SU}(2)$-enhanced topological ${\rm U}(1)$ as ${\rm SU}(2)_R$. The $T[{\rm SU}(2)]$ theory allows us to construct an ${\rm SL}(2,\mathbb{Z})$ transformation acting on the 3-dimensional $\CN=4$ theory $\CT$ that possesses ${\rm SU}(2)_\CT$ flavor symmetry, when the condition for a moment map operator called the \textit{fundamental identity} is satisfied\cite{Gaiotto:2008sd}.\footnote{See (8.1) of \cite{Gaiotto:2008sd}.}  Let's define the modular $S$ transformation as gauging the diagonal part of ${\rm SU}(2)_\CT$ and ${\rm SU}(2)_L$
\begin{align}
    S:\CT\longmapsto\CT_S\equiv \CT\cdot S\equiv\frac{\CT\otimes T[{\rm SU}(2)]}{{\rm SU}(2)^{\rm Diag}}\;.
\end{align}
The resulting theory still possesses ${\rm SU}(2)_{\CT_S}$ identified with the ${\rm SU}(2)_R$ of the $T[{\rm SU}(2)]$ theory used in the transformation. Then with the $T$ transformation defined as adding the supersymmetric Chern-Simons term of unit level associated with ${\rm SU}(2)_\CT$, $S$ and $T$ are known to satisfy the ${\rm SL}(2,\mathbb{Z})$ relation\cite{Gaiotto:2008sd,Terashima:2011qi}
\begin{align}
    S^2=(ST)^3\equiv C\;,\quad C^2=1\;.
\end{align}
Note that after the first $S$ transformation, the fundamental identity is satisfied since the $T[{\rm SU}(2)]$ theory has a nilpotent moment map operator\cite{Gaiotto:2008sd}. See figure~\ref{fig: tsu2 cartoon} and~\ref{fig: su2 SL(2,Z)} below to get the picture.

\begin{center}
    \begin{minipage}[t]{0.25\textwidth}
        \centering
        \begin{minipage}[b]{\textwidth}
            \includegraphics[width=\textwidth]{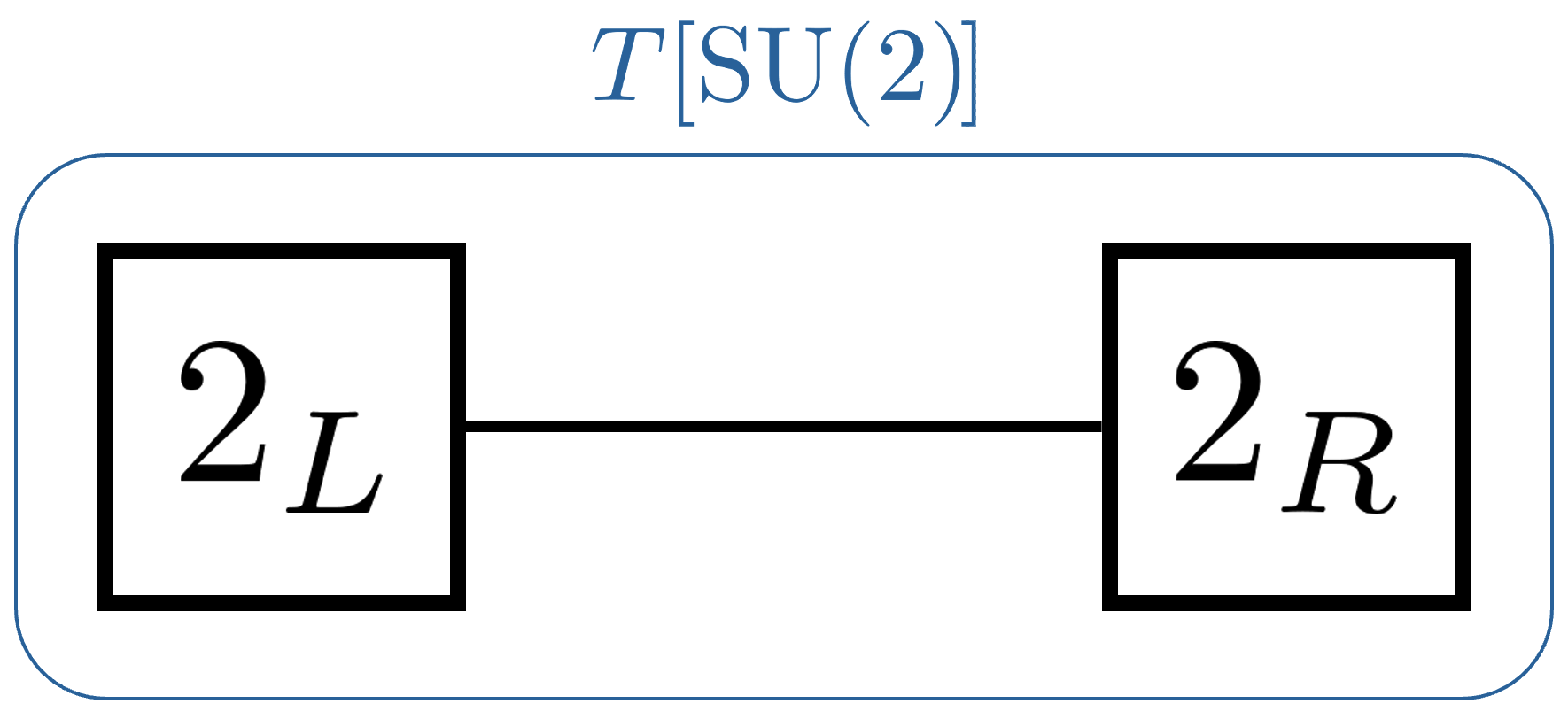}
            \vspace{0.7cm}
        \end{minipage}
        \captionof{figure}{Cartoon of the $T[{\rm SU}(2)]$ theory.}
\label{fig: tsu2 cartoon}
    \end{minipage}
    \hspace{0.3cm}
    \begin{minipage}[t]{0.65\textwidth}
        \centering
        \includegraphics[width=1\textwidth]{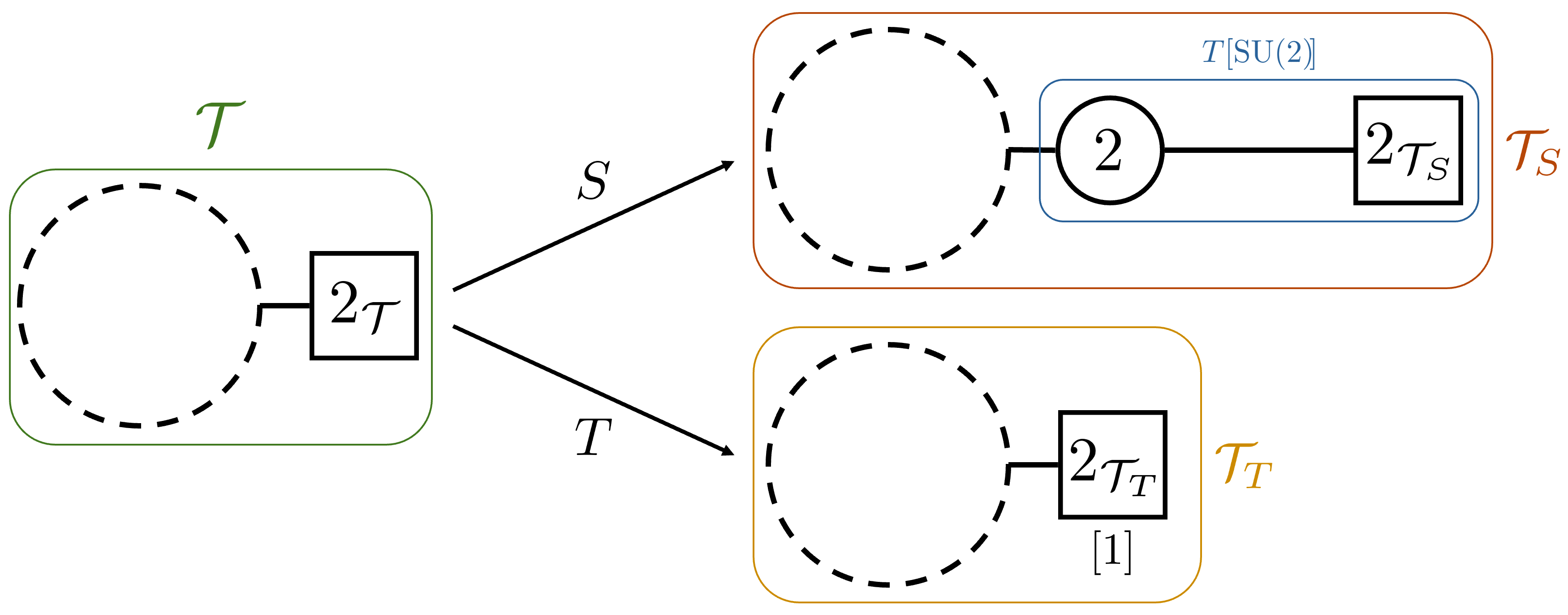}
        \captionof{figure}{$\mathrm{SL}(2,\mathbb{Z})$ action for the ${\rm SU}(2)$ flavor symmetry. $[k]$ under the flavor symmetry box indicates the supersymmetric Chern-Simons term of level $k$.}
\label{fig: su2 SL(2,Z)}
    \end{minipage}
\end{center}

Let's define the $\mathbb{D}(\vec{k}\in\mathbb{N}^{n>1})$\footnote{Compared with previous work \cite{Gang:2024wxz}, we restrict $\vec{k}$ to $\mathbb{N}^{n}$ rather than $(\mathbb{Z}^*)^{n}$.} theory which possesses the two ${\rm SU}(2)$ flavor symmetries ${\rm SU}(2)^1\times {\rm SU}(2)^n$, similar to the $T[{\rm SU}(2)]$ theory to consider generic ${\rm SL}(2,\mathbb{Z})$ transformation. We want it to be a \textit{kernel} theory associated with the ${\rm SL}(2,\mathbb{Z})$ transformation
in a sense that the action gauging the diagonal part of ${\rm SU}(2)_\CT$ and ${\rm SU}(2)^1$ of the $\mathbb{D}(\vec{k})$ theory maps\footnote{${\rm SL}(2,\mathbb{Z})$ actions are acting as right multiplication.}
\begin{align} \label{eq: kernel bD(k)} \begin{split}
    &\varphi(\vec{k})\equiv T^{k_1}ST^{k_2}S\cdots T^{k_{n-1}}ST^{k_n}\in{\rm SL}(2,\mathbb{Z})
    \\&\text{then }\CT\xmapsto{\bigr/\bigr({\rm SU}(2)_\CT={\rm SU}(2)^1\text{ of }\mathbb{D}(\vec{k})\bigr)}\CT_{\varphi(\vec{k})}=\CT\cdot T^{k_1}ST^{k_2}S\cdots T^{k_{n-1}}ST^{k_n}\;.
\end{split}
\end{align}
Through the definition of $S$ and $T$ transformations, we can read the field theory description of the $\mathbb{D}(\vec{k})$ theory from \eqref{eq: kernel bD(k)}
\begin{align}
    {\color{Blue}{\rm SU}(2)^1_{k_1}}+\frac{T[{\rm SU}(2)]^{\otimes (n-1)}}{\color{Orange}{\rm SU}(2)^{2}_{k_2}\otimes {\rm SU}(2)^3_{k_3}\otimes\cdots \otimes {\rm SU}(2)^{n-1}_{k_{n-1}}}+{\color{Blue}{\rm SU}(2)^n_{k_n}}
\end{align}
where ${\rm SU}(2)^{i=1,\cdots,n}$ are defined as
\begin{equation}
\begin{aligned}
    &{\color{Blue}{\rm SU}(2)^1}&:&\,\,{\rm SU}(2)_L\text{ of the first }T[{\rm SU}(2)]\;,
    \\&{\color{Orange}{\rm SU}(2)^{i=2,\cdots,n-1}}&:&\,\, \text{Diagonal subgroup of }({\rm SU}(2)_R\text{ of }(i-1)^\mathrm{th}\,\,T[{\rm SU}(2)])
    \\&&&\times ({\rm SU}(2)_L\text{ of }i^\mathrm{th}\,\,T[{\rm SU}(2)])\;,
    \\&{\color{Blue}{\rm SU}(2)^n}&:&\,\,{\rm SU}(2)_R\text{ of the last }T[{\rm SU}(2)]
\end{aligned}
\end{equation}
and subscripts at each $\color{Orange}{\rm SU}(2)^{i=2,\cdots,n-1}_{k_i}$ indicates adding the supersymmetric Chern-Simons term of level $k_i$ associated with ${\rm SU}(2)^i$ before gauging it. $\color{Blue}{\rm SU}(2)^{1}_{k_{1}}({\rm SU}(2)^n_{k_n})$ indicates adding the supersymmetric Chern-Simons term of level $k_{1}(k_n)$ associated with the flavor $\color{Blue}{\rm SU}(2)^{1}({\rm SU}(2)^n)$.

\begin{figure}[h]
\centering
\includegraphics[width=.70\textwidth]{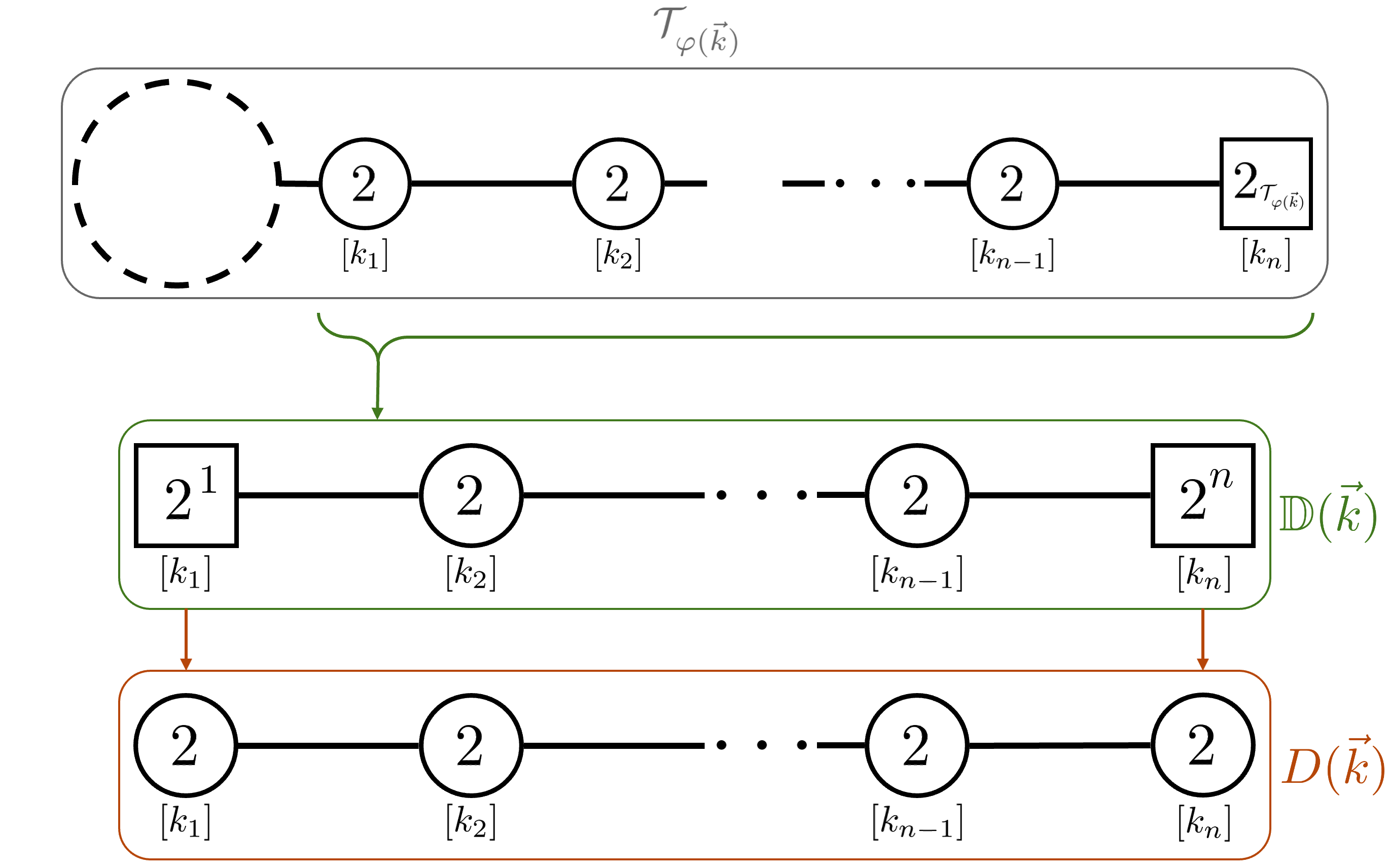}
\captionof{figure}{Construction of the $D(\vec{k})$ theory.}
\label{fig: D(k) construction}
\end{figure}
Now we are ready to define the $D(\vec{k})$ theory we are interested in. After gauging ${\rm SU}(2)^1$ and ${\rm SU}(2)^n$ of the $\mathbb{D}(\vec{k})$ theory, let's fix the real mass parameter associated with ${\rm U}(1)_A$ as zero to ensure the $T[{\rm SU}(2)]$ segments possess $\CN=4$ supersymmetry\cite{Terashima:2011qi}. Then we get the $D(\vec{k})$ theory shown in figure~\ref{fig: D(k) construction}\cite{Gang:2024wxz}
\begin{align}
    D(\vec{k})\equiv \frac{T[{\rm SU}(2)]^{\otimes(n-1)}}{{\rm SU}(2)^1_{k_1}\otimes {\rm SU}(2)^2_{k_2}\otimes\cdots\otimes{\rm SU}(2)^n_{k_n}}\;.
\end{align}
 Note that, due to the presence of Chern-Simons terms, the manifest supersymmetry at the UV is $\CN=3$ rather than $\CN=4$. However, a nilpotency of the moment map operator of the $T[{\rm SU}(2)]$ makes it enhanced to $\CN=4$ in the IR\cite{Gaiotto:2008sd,Gang:2024wxz,Gang:2018huc}. Furthermore, gauging the whole ${\rm SU}(2)^{i=1,2,\cdots,n}$ with $\vec{k}\in\mathbb{N}^n$ is expected to completely lift the Coulomb and Higgs branches of the $T[{\rm SU}(2)]$ segments\cite{Gang:2024wxz}, thereby yielding a rank-0 theory.

\subsection{Field theoretic realization of the WRT TQFT : $D(\vec{k})|_A$ and $\CD(p,q)|_A$ theory} \label{Subsec: D(p,q) theory}
To get the motivation, let's review the ${\rm U}(1)$ Chern-Simons effective description of the fractional quantum Hall effect(FQHE). The ${\rm SL}(2,\mathbb{Z})$ transformations on the 3-dimensional theory possessing ${\rm U}(1)$ flavor symmetry are generated by $S$ and $T$ actions defined as\cite{Witten:2003ya,Burgess:2001}
\begin{align} \label{eq: U(1) SL(2,Z)}
    \begin{split}
        S&:\text{Adding }\CL_{S,{\rm U}(1)}[A_\mu,B_\mu]\equiv\frac{i}{2\pi}\epsilon^{\m\n\r}A_\m \pd_\n B_\r\text{ and makes }A_\m\text{ dynamical}\;,
        \\ T&:\text{Adding Chern-Simons term of unit level}\;.
    \end{split}
\end{align}
Here, $A_\m$ indicates a background vector field associated with the original ${\rm U}(1)$ flavor symmetry. Then starting from the \textit{non-dynamical} ${\rm U}(1)$ Chern-Simons theory of an integer level $k$ which describes a $\nu=k$ integer quantum Hall effect (IQHE)
\begin{align} \label{eq: U(1) IQHE}
    Z_{k}[A_\m]=e^{\frac{ik}{4\pi}\int A\wedge dA}\longrightarrow \frac{\d \log Z}{i\d A_\mu(x)}= \lg j^\m(x)\rg=\frac{k}{4\pi}\e^{\m\n\r}F_{\n\r}\;,\quad \sigma_{xy}\equiv \frac{\n}{2\pi}=\frac{k}{2\pi}\;,
\end{align}
the $S$ transformed theory is the \textit{dynamical} ${\rm U}(1)_k$ Chern-Simons theory and it is known to describe $\nu=-1/k$ FQHE\cite{Witten:2003ya,Burgess:2001}. Combined with the $T$ transformation, which by definition increases $\nu$ by $1$, ${\rm SL}(2,\mathbb{Z})$ transformation results in the Möbius transformation on \(\nu\). Then we can construct the theory that describes the generic fractional number $\nu=-q/p$ FQHE in terms of its \textit{negative continued fraction} expansion
\begin{align}
    -\frac{q}{p}=-\frac{1}{k_1-\frac{1}{k_2-\frac{1}{k_3-\cdots\frac{1}{k_n}}}}\longrightarrow {\rm U}(1)_{k_n}\cdot T^{k_{n-1}}ST^{k_{n-2}}\cdots T^{k_1}S\;.
\end{align}
Focusing on the relation between the level and the conductivity, ${\rm SL}(2,\mathbb{Z})$ transformation allows us to find the Chern-Simons theory of the \textit{fractional level}, by regarding the resulting theory as a ${\rm U}(1)_k$ Chern-Simons theory of an effective level $k=p/q$.

Non-Abelian generalization of \eqref{eq: U(1) SL(2,Z)} is the ${\rm SL}(2,\mathbb{Z})$ transformation we discussed in the previous section~\ref{Subsec: D(k) theory}. Thanks to the supersymmetry, we have the kernel Lagrangian $\CL_{S,{\rm SU}(2)}=\CL_{T[{\rm SU}(2)]}$ to realize the $S$ transformation while the $T$ transformation is defined in the same manner with the ${\rm U}(1)$ case. Now consider the \textit{pure} ${\rm SU}(2)_{k_n-2}$ Chern-Simons theory. We can supersymmetrize it to the $\CN=3$ ${\rm SU}(2)_{k_n}$ Chern-Simons theory from their IR equivalence when $k_n\in\mathbb{N}$\cite{Witten:1999ds,Gang:2024wxz}. Identifying it with the rightmost ${\rm SU}(2)$ in figure~\ref{fig: D(k) construction}, we expect the $D(\vec{k})$ theory to behave as the ${\rm SU}(2)_k$ Chern–Simons theory with a fractional level, based on the intuition gained from the ${\rm U}(1)$ FQHE theory. Therefore, it is natural to identify the $D(\vec{k})$ theory with the $\mathfrak{sl}(2,\mathbb{C})$ WRT TQFT at a generic primitive root of unity parameter $x=e^{\frac{2\pi im}{k+2}}$ since it can be regarded as the ${\rm SU}(2)_{k'}$ Chern-Simons theory with a fractional level $k'$. We can find the modular data of the WRT TQFT up to multiplicity of 2 by simply extending the level of pure ${\rm SU}(2)$ Chern-Simons theory to rational number.

We perform the \textit{topological A twist} introduced in appendix~\ref{Ap: Topl twist} to the $D(\vec{k})$ theory. Resulting $D(\vec{k})|_A$ theory is expected to be a \textit{non-unitary} TQFT\cite{Gang:2021hrd}. Through an anomaly analysis associated with $\mathbb{Z}_2^{\otimes n}$ 1-form symmetry originated from the center subgroups of gauged ${\rm SU}(2)^{i=1,2,\cdots,n}$, we can predict the existence of a decoupled unitary TQFT\cite{Hsin:2018vcg}. We denote it as $\TFTk$. The following property of the $\TFTk$ was studied in the previous work\cite{Gang:2024wxz}:
\begin{align} \label{eq: TFTk prediction}
    \bigr|Z^{S^3}_{\TFTk}\bigr|^{-2}=\underbrace{(\text{The number of Bethe vacua of }\TFTk)}_{=\text{ The dimension of }\TFTk}=\begin{cases}
        2^n&(p\text{ is odd})
        \\2^{n-1}&(p\text{ is even})
    \end{cases}\;.
\end{align}
The remaining non-unitary TQFT sector is predicted to depend only on $p$ and $q$\cite{Gang:2024wxz}. We denote it as $\CD(p,q)|_A$:
\begin{align}
    D(\vec{k})|_A\equiv \CD(p,q)|_A\otimes \TFTk\;.
\end{align}
We claim the first part $\CD(p,q)|_A$ to be the WRT TQFT. In the following sections, we present a variety of evidence.

\subsection{Bethe/Gauge analysis for $D(\vec{k})|_A$ theory}\label{Subsec: D(k) BG analysis}
*The full details of the Bethe/Gauge analysis are provided in appendix~\ref{Ap: BG detail}.

\noindent Consider the squashed 3-sphere\footnote{$S^3_b\equiv\bigr\{x\in\mathbb{R}^4\;\bigr|\;b^2\bigr((x^1)^2+(x^2)^2\bigr)+\frac{1}{b^2}\bigr((x^3)^2+(x^4)^2\bigr)=1\bigr\}\;.$} partition function of the $D(\vec{k})$ theory $Z^{S^3_b}_{D(\vec{k})}[\nu]$. The supersymmetric localization\cite{Pestun:2016zxk} converts the path integral associated with dynamical vector multiplets to the ordinary integral over the real mass parameters which label their BPS configurations and the result depending on the parameters $\nu$ and $b^2$\cite{Hama:2010av,Hama:2011ea,Imamura:2011su,Closset:2018ghr}. $\nu$\footnote{Distinct from the Hall conductivity $\nu$.} is the mixing parameter between ${\rm U}(1)_R$ and ${\rm U}(1)_A$. $b^2$ is the squashing parameter that controls the background metric of the 3-sphere. In our convention, choosing $\nu=-1$ corresponds to performing topological A-twist on the theory. Then we can freely choose the value of $b^2$ to evaluate an integral thanks to the metric independence. A particularly convenient choice is $b^2\rightarrow 0$, which allows us to use the saddle point approximation. From the small $b^2$ expansion of the partition function
\begin{align}
    Z^{S^3_b}_{D(\vec{k})|_A}=\int d\vec{Y}\exp\biggr(\frac{\mathcal{W}_0^{D(\vec{k})|_A}(\vec{Y})}{\hbar}+\mathcal{W}_1^{D(\vec{k})|_A}(\vec{Y})+\mathcal{O}(\hbar)\biggr)\;,\quad \hbar\equiv 2\pi ib^2\;,
\end{align}
the modified saddle point equation(Bethe equation) $e^{\partial \mathcal{W}_0}=1$ defines a set of Bethe vacua\cite{Nekrasov:2014xaa,Gang:2024wxz,Closset:2018ghr}
\begin{align}
    \frac{{\rm BV}^{D(\vec{k})}}{(\textbf{Weyl})}\equiv \;\biggr\{\;V=\vec{Y}_*\;\biggr|\;\exp\bigr(\partial_{\vec{Y}}\mathcal{W}_0^{D(\vec{k})|_A}\bigr)\bigr|_{\vec{Y}=\vec{Y}_*}=1\biggr\}\biggr/\biggr((\textbf{Weyl}),\;V_i\sim V_i+2\pi i\;\biggr)\;.
\end{align}
Here, we introduce the collective notation $\vec{Y}=(\vec{Z},\vec{W})$ for ($2\pi b$-multiplied) real mass parameters $Z_{i=1,2,\cdots,n-1}$ associated with ${\rm U}(1)_G$ of each $T[{\rm SU}(2)]$ segment and $W_{i=1,2,\cdots,n}$ associated with ${\rm SU}(2)^i$. $(\mathbf{Weyl})$ indicates the Weyl subgroup of the gauge group. For each ${\rm SU}(2)^i$, it acts on the Bethe vacuum $V$ as $W_i\longmapsto -W_i$. Equipped with Bethe vacua, we can extract the $\HF$ data\footnote{Note that the definition of $\mathbf{HF}$ data is different from the one in~\cite{Gang:2024wxz}. We defined it with $\CH$ instead of $\CH^{-\frac{1}{2}}$ and we do not divide each $\CF$ by $\CF(V_1)$. From this, the two physically equivalent $\HF$ data can have overall phase difference in $\CF$. However, after determining $V_1$ and dividing each $\CF$ by $\CF(V_1)$, they should be equal due to its physical interpretation\cite{Gang:2021hrd,Wen:2015qwa,Gannon:2003kg}. We denote such relation as $\HF_1\cong \HF_2$.} defined as pairs of Handle-gluing $\CH(V)$ and Fibering $\CF(V)$ for each Bethe vacuum $V$ as
\begin{align} \label{eq: def of HF data}
    \begin{split}
        &\HF^{D(\vec{k})}\equiv\biggr\{\bigr(\CH(V),\CF(V)\bigr)\;\biggr|\;V\in\frac{{\rm BV}^{D(\vec{k})}}{(\mathbf{Weyl})}\biggr\}
        \\&\text{where}
        \\&\CH(V)\equiv \frac{e^{i\d}}{|\textbf{Weyl}|^2}\det\left[-\partial^2_{\vec{Y}}\CW_0^{D(\vec{k})|_A}\right]e^{-2\CW_1^{D(\vec{k})|_A}}\biggr|_{V},\quad\CF(V)\equiv e^{-\frac{1-\vec{Y}\cdot\pd_{\vec{Y}}}{2\pi i}\CW^{D(\vec{k})|_A}_0}\biggr|_V\;.
    \end{split}
\end{align}
Here, $|\mathbf{Weyl}|$ indicates the dimension of the Weyl subgroup of the gauge group, $2^n$ for our case. The overall phase factor $e^{i\delta}$ can be determined from the condition
\begin{align}
    \sum_{V\in\frac{{\rm BV}^{D(\vec{k})}}{(\textbf{Weyl})}}\CH^{-1}(V)=1\;.
\end{align}
However, if we strictly set the value of $\n$ to $-1$, the contributions of the adjoint chirals in the $T[{\rm SU}(2)]$ segments to $\CW_1^{D(\vec{k})}$ become divergent, making the Handle-gluing operators singular. As a prescription, we consider the double-scaling ansatz by introducing a small parameter $\e$:
\begin{align}
    \n=-1+\e\;,\quad Z_i=(-1)^{G_i}W_i+f_i\e+\CO(\e^2)\;,\quad G_i\in \mathbb{Z}_2\text{ for }i=1,2,\cdots,n-1
\end{align}
and define observables in $\e\rightarrow 0$ limit. Substituting ansatz into un-exponentiated Bethe equation $\pd \CW_0\in 2\pi i\mathbb{Z}$, we get the matrix equation $\mathrm{K}\vec{W}^{\vec{G}} \in \pi i \mathbb{Z}^n$ where
\begin{align}
        \mathrm{K}[\vec{k}]\equiv \begin{pmatrix}
        k_1 & 1 & 0 & 0&\cdots &0&0&0 \\
        1 & k_2 & 1 & 0 & \cdots&0&0&0\\
        0&1&k_3  & 1 & \cdots &0&0&0\\
        0 &0 & 1 & k_4  & \cdots&0&0&0\\
        \vdots&\vdots&\vdots&\vdots&\ddots&\vdots&\vdots&\vdots\\
        0&0&0&0&\cdots&k_{n-2}&1&0\\
        0&0&0&0&\cdots&1&k_{n-1}&1\\
        0&0&0&0&\cdots&0&1&k_n\\    
    \end{pmatrix}\text{ and }W_i^{\vec{G}}\equiv W_i\prod_{j=1}^{i-1}(-1)^{G_j}
\end{align}
with the consistency equation which determines $f_i$ from the solution $\vec{W}$
\begin{align} \label{eq: Consist eqn}
    e^{2W_{i+1}}\frac{\pi i+2f_i}{\pi i-2f_i}=-1\text{ for }i=1,2,\cdots,n-1\;.
\end{align}
Note that \eqref{eq: Consist eqn} \textit{forbids} $W_{i+1}\in\pi i\mathbb{Z}$. We find the following set of solutions labeled by $(|p|-1)\times 2^{n-1}\times 2^{n}$ parameters $\a\in\{1,2,\cdots,|p|-1\}$, $\vec{n}\in(\mathbb{Z}^{n-1}_2,0)$ and $\vec{G}\in\mathbb{Z}_2^n$
\begin{align} \label{eq: D(k) Before WQ BV}
    {\rm BV}^{D(\vec{k})}\equiv \biggr\{\bigr(Z_i=(-1)^{G_i}W_i^{\vec{n},\vec{G}},W_i^{\vec{n},\vec{G}}\bigr)\;\biggr|\;W_i^{\vec{n},\vec{G}}=\pi i\biggr(n_i+\frac{q_i}{p}\a\biggr)(-1)^{G_n + \sum_{j=1}^{i-1}G_j}\biggr\}
\end{align}
where $p_i$ and $q_i$ are defined as
\begin{align}
    p_i\equiv \det(\mathrm{K}[k_i,\cdots,k_n])\;,\quad\frac{q_i}{p_i} \equiv \frac{(-1)^{i+1}}{k_i - \frac{1}{k_{i+1}-\frac{1}{k_{i+2}-\cdots \frac{1}{k_n}}}}\;,\quad p\equiv p_1\;,\quad q\equiv q_1\;.
\end{align}
Since the elements that differ only in $\vec{G}$ are connected by the sign difference especially in $W_i$\footnote{Also note that the Handle-gluing and the Fibering are independent of the choice of $\vec{G}$.}, we take the Weyl-quotient by fixing $\vec{G}=\vec{0}$ to get $(|p|-1)\times 2^{n-1}$ Bethe vacua
\begin{align} \label{eq: D(k) WQ Bethe vacuum}
    \frac{{\rm BV}^{D(\vec{k})}}{(\mathbf{Weyl})}={\rm BV}^{D(\vec{k})}\biggr|_{\vec{G}=\vec{0}}=\biggr\{V(\a,\vec{n})\equiv \bigr(Z_i=W^{\vec{n}}_i,W^{\vec{n}}_i\bigr)\;\biggr|\;W^{\vec{n}}_i=\pi i\biggr(n_i+\frac{q_i}{p}\a\biggr)\biggr\}\;.
\end{align}
For each Bethe vacuum labeled by $\a\in\{1,2,\cdots,|p|-1\}$ and $\vec{n}\in(\mathbb{Z}_2^{n-1},0)$, the Handle-gluing and the Fibering are evaluated as
\begin{align} \label{eq: HF data of D(k)}
\begin{split}
    \CH(\a,\vec{n})&=2^{n-2}|p|\sin^{-2}\biggr(\frac{\pi}{p}\a\biggr)\quad(\text{Independent of }\vec{n})\;,
    \\\CF(\a,\vec{n})&=\CF_0\exp\biggr(\frac{\pi i}{2p}q\a^2+\pi in_1\a+\frac{\pi i}{2}\vec{n}^T{\rm K}[\vec{k}]\vec{n}\biggr)\quad (\CF_0\equiv e^{\frac{5}{12}(n-1)\pi i})\;.
\end{split}
\end{align}

\paragraph{Comment about excluded cases} We \textit{assume} that there is no $W_{2,3,\cdots,n}\in\pi i\mathbb{Z}$ and every $(\a,\vec{n},\vec{G})$ corresponds to a distinct pair $(e^{\vec{Z}},e^{\vec{W}})$. This is not true in general. We observed problematic cases arise when $n\ge 3$. Regardless of whether $p$ is even or odd, some choices of $\vec{k}$ yield $W_{2,3,\cdots,n}\in\pi i\mathbb{Z}$. We also checked for the following choices of $\vec{k}$ correspond to even $p$
\begin{align}
    \vec{k}=\text{(even,odd,even), (even,even,odd,odd), (even,odd,even,odd,even), }\cdots
\end{align}
redundant elements are induced and the total number of ${\rm BV}^{D(\vec{k})}$ is reduced to some subtle number, which is not a divisor of $(|p|-1)\times 2^{n}\times 2^{n-1}$. We will exclude such cases. For example,
\begin{align}
    \vec{k}=(3,5,3),\;(5,2,3,2),\;(2,3,2,2,3),\;\cdots
\end{align}
for odd $p$ and
\begin{align}
    \vec{k}=(4,3,2),\;(2,2,5,3),\;(3,2,2,2,3),\;\cdots
\end{align}
for even $p$.

\paragraph{3-sphere partition function} Let's define $m_0^\pm\in\{1,2,\cdots,|p|-1\}$ such that
\begin{align}
    |q|m_0^\pm\in |p|\mathbb{N}_0\pm 1\quad(m_0^-=|p|-m_0^+)\;.
\end{align}
We can evaluate the absolute value of the 3-sphere partition function from the $\mathbf{HF}$ data of the $D(\vec{k})|_A$ theory \eqref{eq: HF data of D(k)} as\cite{Closset:2018ghr}
\begin{align}
    \bigr|Z^{S^3}_{D(\vec{k})|_A}\bigr|=\biggr|\sum_{\HF^{D(\vec{k})}}\CH^{-1}\CF\biggr|=\sqrt{\frac{1}{2^{n-2}|p|}}\sin\biggr(\frac{\pi}{|p|}m_0^\pm\biggr)\;.
\end{align}
Note that the set of Bethe vacua is larger than the set of saddle points used in the method of steepest descent in general since it allows $\partial_{\vec{Y}}\mathcal{W}_0\in2\pi i\vec{\mathbb{Z}}\setminus\{\vec{0}\}$. It is conjectured that we can find the one special \textit{true Bethe vacuum} $V_1$ such that we can construct the steepest descent contour passing through it to localize the integral around the $V_1$. It implies
\begin{align}
\label{eq: True vac condition}
    \bigr|Z^{S^3}_{D(\vec{k})|_A}\bigr|=\bigr|\CH(V_1)\bigr|^{-\frac{1}{2}}
\end{align}
and the phase of the integral will be determined from the choice of the contour. See \cite{Gang:2025} for concrete examples. For the $D(\vec{k})|_A$ theory, we find the candidates
\begin{align} \label{eq: True vac of D(k)}
    V_1\in\bigr\{V(\a,\vec{n})\;\bigr|\;\a\in\{m_0^+,m_0^-\},\;\vec{n}\in(\mathbb{Z}^{n-1}_2,0)\bigr\}\;.
\end{align}

\section{Matching the field theory with the WRT TQFT} \label{Sec: WRT data extraction}
This section contains the main results of the paper. We assert the following correspondence
\begin{align}
    \begin{split}
        &\WRT^m_k\text{ : }\text{WRT TQFT constructed from the primitive root of unity }x=e^{\frac{2\pi im}{k+2}}
        \\&=\CD(p,q)|_A\subset D(\vec{k})|_A
    \end{split}
\end{align}
with concrete relation between $(k,m)$ and $(p,q)$. Final results are summarized in table~\ref{tab: Results summary}. See appendix~\ref{Ap: WRT review} for the exact definition and modular data of $\WRT^m_k$.
\begin{table}[h]
    \centering
    \renewcommand{\arraystretch}{1.2} 
    \begin{tabular}{ll l lll}
    \toprule
    \multicolumn{2}{c}{\textbf{Odd} $\boldsymbol{p}$} & & \multicolumn{3}{c}{\textbf{Even} $\boldsymbol{p}$} \\
    \cmidrule(r){1-2} \cmidrule(l){4-6} 
    
    $s\times \mathrm{sgn}(pq)$ & Identified TQFT & & $s\times \mathrm{sgn}(pq)$ & $(-1)^N$ & Identified TQFT \\ 
    \midrule
    
    \multirow{2}{*}{$+1$} & \multirow{2}{*}{$\mathrm{WRT}_{|p|-2}^{m}$} & & \multirow{2}{*}{$+1$} & $+1$ & $\mathrm{WRT}_{|p|-2}^{m}$ \\
     & & & & $-1$ & $\mathrm{WRT}_{|p|-2}^{2|p|+m}$ \\ 
    \addlinespace 
    
    \multirow{2}{*}{$-1$} & \multirow{2}{*}{$\mathrm{WRT}_{|p|-2}^{2|p|-m}$} & & \multirow{2}{*}{$-1$} & $+1$ & $\mathrm{WRT}_{|p|-2}^{4|p|-m}$ \\
     & & & & $-1$ & $\mathrm{WRT}_{|p|-2}^{2|p|-m}$ \\ 
    \bottomrule
    \end{tabular}
    \captionof{table}{Summary of identified WRT TQFTs.
    For \textit{odd} \(p\), there are unique even $m\in\{2,4,6,\cdots,|p|-1\}$ and $s\in\{\pm1\}$ satisfy $|q|m\in|p|\mathbb{N}_0+s$. 
    For \textit{even} \(p\), there are unique odd $m\in\{1,3,5,\cdots,|p|-1\}$, $N\in\mathbb{N}_0$ and $s\in\{\pm1\}$ satisfy $|q|m=2|p|N+s$. Identification can be done by $m$, $N$ and $s$.}
    \label{tab: Results summary}
\end{table}

$\mathbf{HF}$ data of the $D(\vec{k})|_A$ theory introduced in the previous section~\ref{Subsec: D(k) BG analysis} (See \eqref{eq: HF data of D(k)}) is always factorized into the two reasonable $\mathbf{HF}$ data
\begin{align} \label{eq: HF factorization}
    \HF^{D(\vec{k})}=\HF^{\CD(p,q)}\otimes\HF^{{\rm TFT}[\vec{k}]}
\end{align}
in a sense that each of them satisfies the characteristic property
\begin{align} \label{eq: HF data relation}
    \bigr(\bigr|Z^{S^3_b}\bigr|=\bigr)\biggr|\sum_{\HF}\CH^{-1}\CF\biggr|=\bigr|\CH_1\bigr|^{-\frac{1}{2}}\text{ for some }\CH_1\in\HF
\end{align}
and $\HF^{\TFTk}$ is in agreement with the prediction \eqref{eq: TFTk prediction}. We identified $\HF^{\CD(p,q)}$ with $\HF$ data of the WRT TQFT.

\textit{When $p$ is odd}, $(k,m)$ is uniquely determined from $(p,q)$ and the stronger correspondence can be made by finding the full modular matrices of the $D(\vec{k})|_A$ theory. We find various candidates for simple objects that allow us to construct modular $S$ and $T$ matrices. The resulting matrices are factorized
\begin{align}
    S^{D(\vec{k})}=S^{\CD(p,q)}\otimes S^{\TFTk;\vec{L}}\;,\quad T^{D(\vec{k})}=T^{\CD(p,q)}\otimes T^{\TFTk;\vec{L}}
\end{align}
and we identify the $\CD(p,q)$ part with modular matrices of the WRT TQFT, determined through the $\HF$ data analysis. The other part is modular matrices of the decoupled TQFT. We have an ambiguity in identifying the correct maximally independent set of simple objects and the label $\vec{L}\in\mathbb{Z}^{n-1}_2$ is introduced to enumerate all possibilities.

\textit{When $p$ is even}, $(k,m)$ is NOT uniquely determined from $(p,q)$ through the $\HF$ data analysis. We also find difficulties in constructing full modular matrices. However, we can construct submatrices which can be interpreted as modular matrices of the $\CD(p,q)|_A$ theory in the vacuum sector of the decoupled $\TFTk$. We identify them with modular matrices of the WRT TQFT. During the procedure, we can uniquely determine $(k,m)$ from $(p,q)$.

\subsection{HF data analysis} \label{Subsec: HF data analysis}
\paragraph{When $p$ is odd} We find the factorization \eqref{eq: HF factorization} as
\begin{align}
\label{eq: odd p HF factorization}
\begin{split}
        \mathbf{HF}^{\mathcal{D}(p,q)}&=\biggr\{\biggr(\frac{|p|}{4}\sin^{-2}\bigr(\frac{2\pi a}{p}\bigr),\mathcal{F}_0\exp\bigr(\frac{2\pi i}{p}q a^2\bigr)\biggr)\;\biggr|\;a\in\bigr\{1,2,\cdots,\frac{|p|-1}{2}\bigr\}\biggr\}\;,
        \\\mathbf{HF}^{\TFTk}&= \biggr\{\biggr(2^{n},e^{\pi i(\frac{pq}{2}+n_1)t}\exp\bigr(\frac{\pi i}{2}\vec{n}^T \mathrm{K}[\vec{k}]\vec{n}\bigr)\biggr)\;\biggr|\;t\in\mathbb{Z}_2,\;\vec{n}\in(\mathbb{Z}^{n-1}_2,0)\biggr\}\;.
\end{split}
\end{align}
There is the unique even $m\in\{2,4,6,\cdots,|p|-1\}$ and $s\in\{\pm 1\}$ satisfy
\begin{align}
    |q|m\in |p|\mathbb{N}_0+s\;.
\end{align}
When $s\times {\rm sgn}(pq)>0$, $\HF^{\mathcal{D}(p,q)}$ is identified with $\HF$ data of $\WRT^{m}_{|p|-2}$ up to an overall phase difference in the Fibering data:
\begin{align}
    \begin{split}
        \HF^{\mathcal{D}(p,q)}&\cong \HF^{\WRT^m_{|p|-2}}
        \\&\cong\biggr\{\biggr(\frac{|p|}{4}\sin^{-2}\bigr(\frac{m\pi}{|p|}(2a-1)\bigr),e^{2\pi im\frac{a(a-1)}{|p|}}\biggr)\;\biggr|\;a\in\bigr\{1,2,\cdots,\frac{|p|-1}{2}\bigr\}\biggr\}\;.
    \end{split}
\end{align}
When $s\times {\rm sgn}(pq)<0$, $\HF^{\mathcal{D}(p,q)}$ is identified with $\HF$ data of $\WRT^{2|p|-m}_{|p|-2}$ up to an overall phase difference in the Fibering data:
\begin{align}
    \begin{split}
        \HF^{\mathcal{D}(p,q)}&\cong \HF^{\WRT^{2|p|-m}_{|p|-2}}
        \\&\cong\biggr\{\biggr(\frac{|p|}{4}\sin^{-2}\bigr(\frac{m\pi}{|p|}(2a-1)\bigr),e^{-2\pi im\frac{a(a-1)}{|p|}}\biggr)\;\biggr|\;a\in\bigr\{1,2,\cdots,\frac{|p|-1}{2}\bigr\}\biggr\}\;.
    \end{split}
\end{align}
For both cases, we find the unique $a=m/2$ element in $\HF^{\mathcal{D}(p,q)}$ satisfies \eqref{eq: HF data relation}. Using $m$, we can rewrite the true vacuum of the $D(\vec{k})|_A$ theory \eqref{eq: True vac of D(k)} as
\begin{align} \label{eq: odd p possible V1 in sec3}
    V_1=V\bigr(|p|t_0+(-1)^{t_0}m,\vec{n}_0\bigr)\in\frac{{\rm BV}^{D(\vec{k})}}{(\textbf{Weyl})}\;,
\end{align}
labeled by $t_0\in\mathbb{Z}_2$ and $\vec{n}_0\in(\mathbb{Z}^{n-1}_2,0)$.

For the other sector, we can check $\HF^{\TFTk}$ is consistent with prediction \eqref{eq: TFTk prediction} since the 3-sphere partition function is evaluated as
\begin{align}
    \bigr|Z^{S^3}_{\TFTk}\bigr|=\biggr|\sum_{\HF^{\TFTk}}\CH^{-1}\CF\biggr|=2^{-\frac{n}{2}}
\end{align}
as expected. However, we find no new information about the true vacuum, since all of the Handle-gluings are identical to $2^n$. This fact is reflected in \eqref{eq: odd p possible V1 in sec3} through the ambiguity labeled by $t_0$ and $\vec{n}_0$, whose degree of freedom is the same as the expected dimension of the $\TFTk$.

\paragraph{When $p$ is even} We find the factorization \eqref{eq: HF factorization} as
\begin{align}
\label{eq: even p HF factorization}
\begin{split}
        \mathbf{HF}^{\mathcal{D}(p,q)}&=\biggr\{\biggr(\frac{|p|}{2}\sin^{-2}\bigr(\frac{\pi \a}{p}\bigr),\mathcal{F}_0\exp\bigr(\frac{\pi i}{2p}q \a^2\bigr)\biggr)\;\biggr|\;\a\in\{1,2,\cdots,|p|-1\}\biggr\}\;,
        \\\mathbf{HF}^{\TFTk}&= \biggr\{\biggr(2^{n-1},e^{-\pi i\frac{pq}{2}n_1}\exp\bigr(\frac{\pi i}{2}\vec{n}^T \mathrm{K}[\vec{k}]\vec{n}\bigr)\biggr)\;\biggr|\;\vec{n}\in(\mathbb{Z}^{n-1}_2,0)\biggr\}\;.
\end{split}
\end{align}
There is the unique odd $m\in\{1,3,5,\cdots,|p|-1\}$, $N\in\mathbb{N}_0$ and $s\in\{\pm1\}$ satisfy
\begin{align}
    |q|m=2|p|N+s\;.
\end{align}
When $s\times{\rm sgn}(pq)>0$, $\HF^{\CD(p,q)}$ is identified with $\HF$ data of ${\rm WRT}^m_{|p|-2}$ \textit{and} ${\rm WRT}^{2|p|+m}_{|p|-2}$ up to an overall phase difference in the Fibering data
\begin{align}
    \begin{split}
        \HF^{\CD(p,q)}&\cong \HF^{{\rm WRT}^m_{|p|-2}}\cong\biggr\{\biggr(\frac{|p|}{2}\sin^{-2}\bigr(\frac{m\pi}{|p|}\a\bigr),e^{\frac{\pi im\a^2}{2|p|}}\biggr)\;\biggr|\;\a\in\{1,2,\cdots,|p|-1\}\biggr\}
        \\\text{and}\qquad\;
        \\\HF^{\CD(p,q)}&\cong \HF^{{\rm WRT}^{2|p|+m}_{|p|-2}}\cong\biggr\{\biggr(\frac{|p|}{2}\sin^{-2}\bigr(\frac{m\pi}{|p|}\a\bigr),e^{\frac{\pi im\a^2}{2|p|}+\pi i\a}\biggr)\;\biggr|\;\a\in\{1,2,\cdots,|p|-1\}\biggr\}\;.
    \end{split}
\end{align}
When $s\times {\rm sgn}(pq)<0$, $\HF^{\CD(p,q)}$ is identified with $\HF$ data of ${\rm WRT}^{4|p|-m}_{|p|-2}$ \textit{and} ${\rm WRT}^{2|p|-m}_{|p|-2}$ up to an overall phase difference in the Fibering data
\begin{align}
    \begin{split}
        \HF^{\CD(p,q)}&\cong \HF^{{\rm WRT}^{4|p|-m}_{|p|-2}}\cong\biggr\{\biggr(\frac{|p|}{2}\sin^{-2}\bigr(\frac{m\pi}{|p|}\a\bigr),e^{\frac{-\pi im\a^2}{2|p|}}\biggr)\;\biggr|\;\a\in\{1,2,\cdots,|p|-1\}\biggr\}
        \\\text{and}\qquad\;
        \\\HF^{\CD(p,q)}&\cong \HF^{{\rm WRT}^{2|p|-m}_{|p|-2}}\cong\biggr\{\biggr(\frac{|p|}{2}\sin^{-2}\bigr(\frac{m\pi}{|p|}\a\bigr),e^{\frac{-\pi im\a^2}{2|p|}-\pi i\a}\biggr)\;\biggr|\;\a\in\{1,2,\cdots,|p|-1\}\biggr\}\;.
    \end{split}
\end{align}
In contrast to the previous $p={\rm odd}$ case, we cannot uniquely determine the identified WRT TQFT. We also cannot uniquely determine the element of $\HF^{\CD(p,q)}$ corresponding to the true vacuum since we have the two $\a=m$ and $\a=|p|-m$ elements satisfying \eqref{eq: HF data relation}. For the other sector, we can check $\HF^{\TFTk}$ is consistent with prediction \eqref{eq: TFTk prediction} since the 3-sphere partition function is evaluated as
\begin{align}
    \bigr|Z^{S^3}_{\TFTk}\bigr|=\biggr|\sum_{\HF^{\TFTk}}\CH^{-1}\CF\biggr|=2^{-\frac{n-1}{2}}
\end{align}
as expected. However, we find no new information about the true vacuum, again, since all of the Handle-gluings are identical to $2^{n-1}$. The true vacuum ambiguity of the $D(\vec{k})|_A$ theory remains as $2^n$, which is \textit{larger} than the expected dimension $2^{n-1}$ of the $\TFTk$.

\subsection{Full modular matrix analysis}
Using the method introduced in appendix~\ref{Ap: Method full modular matrix}\cite{Gang:2021hrd}, we can construct full modular $S$ and $T$ matrices of the $D(\vec{k})|_A$ theory from the simple objects summarized in appendix~\ref{Ap: Simple obejcts}. In brief, the procedure is as follows. Let's label simple objects of the $D(\vec{k})|_A$ as
\begin{align}
    \CO[A=1,2,\cdots,(|p|-1)\times2^{n-1}]\;,\quad \CO[1]\equiv 1(\text{Triv.})\;.
\end{align}
Suppose we correctly map Bethe vacua and simple objects(See around \eqref{eq: BVSO map})
\begin{align}
    V_{A=1,2,\cdots,(|p|-1)\times 2^{n-1}}\in\frac{{\rm BV}^{D(\vec{k})}}{(\textbf{Weyl})}\longleftrightarrow \CO[A]\;.
\end{align}
Then the full modular matrices are determined as
\begin{align}
    S^{D(\vec{k})}_{A'A}=\CO[A]\bigr|_{V_{A'}}S^{D(\vec{k})}_{1A'}\;,\quad T^{D(\vec{k})}_{A'A}=\xi \d_{A'A}\CF(V_{A'})
\end{align}
and they should satisfy various physical conditions introduced after \eqref{eq: S matrix condition starting point}. Surprisingly, it almost uniquely determines both the map $V_A$ and modular matrices. However, since calculations are quite lengthy, we present results only in this section. See appendix~\ref{Ap: Detail of S matrix computation} and~\ref{Ap: Explicit TFTk} for details.

\subsubsection{When $p$ is odd}
We have $2^{n-1}$ ambiguity in identifying the correct maximally independent set of simple objects, labeled by $\vec{L}\in\mathbb{Z}^{n-1}_2$. See \eqref{eq: odd p alt Simples} for the definition of $\vec{L}$. For each choice of $\vec{L}$, modular matrices of the $D(\vec{k})|_A$ theory are factorized as
\begin{align}
    S^{D(\vec{k})}=S^{\CD(p,q)}\otimes S^{\TFTk;\vec{L}}\;,\quad T^{D(\vec{k})}=T^{\CD(p,q)}\otimes T^{\TFTk;\vec{L}}
\end{align}
and $S^{\CD(p,q)}$, $T^{\CD(p,q)}$ are identified with modular matrices of the WRT TQFT which we determined its parameters in section~\ref{Subsec: HF data analysis}. With a unique even $m\in\{2,4,6,\cdots,|p|-1\}$ and $s\in\{\pm 1\}$ satisfy $|q|m\in|p|\mathbb{N}_0+s$, we find
\begin{align}
\begin{split}
    &S^{\CD(p,q)}_{ab}=S^{{\rm WRT}^{m,2|p|-m}_{|p|-2}}_{ab}=\e\sqrt{\frac{4}{|p|}}\sin\bigr(\frac{m\pi(2a-1)(2b-1)}{|p|}\bigr)
    \\&\text{with}
    \\&\begin{cases}
        T_{ab}^{\CD(p,q)}=T^{{\rm WRT}^m_{|p|-2}}_{ab}=\xi \d_{ab}\exp\bigr(\frac{2\pi ima(a-1)}{|p|}\bigr)&(s\times{\rm sgn}(pq)>0)
        \\T_{ab}^{\CD(p,q)}=T^{{\rm WRT}^{2|p|-m}_{|p|-2}}_{ab}=\xi \d_{ab}\exp\bigr(-\frac{2\pi ima(a-1)}{|p|}\bigr)&(s\times{\rm sgn}(pq)<0)
    \end{cases}\;.
\end{split}
\end{align}
Here, $a,b\in\bigr\{1,2,\cdots,\frac{|p|-1}{2}\bigr\}$ and $\e\in\{\pm 1\}$ is the sign factor defined as \eqref{eq: Sign factor for WRT S} which guarantees an existence of the positive row in a modular $S$ matrix. $\xi$ can be determined up to $e^{\frac{2\pi i}{3}\mathbb{Z}}$ factor, from the ${\rm SL}(2,\mathbb{Z})$ relation $S^2=(ST)^3$. 

The other $\TFTk$ part becomes non-trivial large modular matrices of size $2^n\times 2^n$. See appendix~\ref{Ap: Explicit TFTk}. During the determination of modular matrices of the $\TFTk$, the true vacuum of the $D(\vec{k})|_A$ theory $V_1$ is uniquely determined as
\begin{align}
    V_1=V(|p|-m,\vec{n}_0)\text{ where }(\vec{n}_0)_{i=1,2,\cdots,n-1}\equiv\frac{1}{2}\bigr(1+(-1)^{q_i+L_i}\bigr),\;\vec{n}_0\in(\mathbb{Z}_2^{n-1},0)\;.
\end{align}
However, it has an ambiguity labeled by $\vec{L}\in\mathbb{Z}^{n-1}_2$ comes from the fact that we don't know which $\vec{L}$ corresponds to the correct set of simple objects of the $\TFTk$.

\subsubsection{When $p$ is even}
We expect that \eqref{eq: even p Simples} cannot be a correct maximally independent set of simple objects of the $D(\vec{k})|_A$ theory. Here is the argument. After constructing the full modular $S$ matrix using a set of Wilson lines \eqref{eq: even p Simples} with undetermined parameter $\vec{Q}$ labels charges of $\mathrm{U}(1)$ lines, let's require the Weyl invariance:
\begin{center}
    $S^{\TFTk}$\emph{ should be invariant under the Weyl transformation of Bethe vacua}
    \begin{align}
        S^{\TFTk}\xrightarrow{(\textbf{Weyl})}S^{\TFTk}
    \end{align}
\end{center}
while assuming that the Bethe vacuum-Simple object map remains unchanged. It leads us to conclude that there is no such $\vec{Q}$ which makes \eqref{eq: even p Simples} as the possible maximally independent set of simple objects of the $D(\vec{k})|_A$ theory which allows us to construct full modular matrices. The best we can do is construct modular matrices of the $\CD(p,q)|_A$ theory embedded in the vacuum sector of the $\TFTk$. With a unique odd $m\in\{1,2,\cdots,|p|-1\}$, $N\in\mathbb{N}_0$ and $s\in\{\pm 1\}$ satisfy $|q|m=2|p|N+s$, we find
\begin{align}
    \begin{split}
        &S^{\CD(p,q)}_{\a\b}=S^{{\rm WRT}^{m,2|p|\pm m,4|p|-m}_{|p|-2}}_{\a\b}=\e\sqrt{\frac{2}{|p|}}\sin\bigr(\frac{m\pi\a\b}{|p|}\bigr)
        \\&\text{with}
        \\&\begin{cases}
            T^{\CD(p,q)}_{\a\b}=T^{{\rm WRT}^{m}_{|p|-2}}_{\a\b}=\xi\d_{\a\b}\exp\bigr(\frac{\pi im\a^2}{2|p|}\bigr) &(s\times {\rm sgn}(pq)>0\text{ and }(-1)^{N}>0)
            \\T^{\CD(p,q)}_{\a\b}=T^{{\rm WRT}^{2|p|+m}_{|p|-2}}_{\a\b}=\xi\d_{\a\b}\exp\bigr(\frac{\pi im\a^2}{2|p|}+i\pi\a\bigr) &(s\times {\rm sgn}(pq)>0\text{ and }(-1)^{N}<0)
            \\T^{\CD(p,q)}_{\a\b}=T^{{\rm WRT}^{4|p|-m}_{|p|-2}}_{\a\b}=\xi\d_{\a\b}\exp\bigr(-\frac{\pi im\a^2}{2|p|}\bigr) &(s\times {\rm sgn}(pq)<0\text{ and }(-1)^{N}>0)
            \\T^{\CD(p,q)}_{\a\b}=T^{{\rm WRT}^{2|p|-m}_{|p|-2}}_{\a\b}=\xi\d_{\a\b}\exp\bigr(-\frac{\pi im\a^2}{2|p|}-i\pi\a\bigr) &(s\times {\rm sgn}(pq)<0\text{ and }(-1)^{N}<0)
        \end{cases}\;.
    \end{split}
\end{align}
Here, $\a,\b\in\{1,2,\cdots,|p|-1\}$ and $\e\in\{\pm 1\}$ is the sign factor defined as \eqref{eq: Sign factor for WRT S} which guarantees an existence of the positive row in a modular $S$ matrix. $\xi$ can be determined up to $e^{\frac{2\pi i}{3}\mathbb{Z}}$ factor, from the ${\rm SL}(2,\mathbb{Z})$ relation $S^2=(ST)^3$.

During the derivation, the true vacuum of the $D(\vec{k})|_A$ theory $V_1$ is restricted as
\begin{align}
    V_1=V(m,\vec{n}_0)\in\frac{{\rm BV}^{D(\vec{k})}}{(\textbf{Weyl})}\;,\quad \vec{n}_0\in(\mathbb{Z}^{n-1}_2,0)
\end{align}
and now the size of an ambiguity becomes the same with the dimension of the $\TFTk$. This is consistent with our ignorance about the $\TFTk$.

\section{Discussion} \label{Sec: Discussion}
In this paper, we studied the $D(\vec{k})$ theory, which consists of an arbitrary number of $T[\mathrm{SU}(2)]$ segments, at the A-twist point, as a generalization of the previous work \cite{Gang:2024wxz}.

\paragraph{Refined statement for an IR equivalence} We can provide a refined version of the prediction made in the previous work \cite{Gang:2024wxz}, which states that
\begin{itemize}
    \item $\CD(p,q)$ is independent of the choice of $\vec{k}$ for given $(p,q)$.
    \item $\CD(p,q)$ and $\CD(p,q+p\mathbb{Z})$ are IR equivalent up to \textit{minimal topological operations} which preserve the absolute value of partition functions.
\end{itemize}
For the first one, we have already provided evidence in section~\ref{Sec: WRT data extraction}, by clarifying modular data of the non-unitary TQFT $\mathcal{D}(p,q)|_A$ in terms that depend only on $p$ and $q$. For the second one, we can verify that it holds when \textit{$p$ is odd}, whereas the modular data exhibit a periodicity of $4$ when \textit{$p$ is even}.

\paragraph{On the decoupled $\TFTk$}
It seems clear that the $D(\vec{k})|_A$ is the possible field theoretic realization of the non-unitary $\mathfrak{sl}(2,\mathbb{C})$ WRT TQFT (up to a decoupled TQFT), but it is still unclear what $\TFTk$ is. The biggest difficulty comes from the fact that we don't know which set of line operators at the UV becomes the correct set of simple objects in the IR. For the case when $p$ is odd, it leads us to the true vacuum ambiguity labeled by the choice of $\vec{L}\in\mathbb{Z}_2^{n-1}$ which obstructs us in uniquely determining the full modular matrices. For the case when $p$ is even, it seems there is no possible candidate for the maximally independent set of simple objects if we assume its form as the tensor product of Wilson lines. However, there are many more line operators at the UV we've not considered. For the simplest possibility, the sum of two Wilson lines can be a simple object. See discussion around (5.3) of~\cite{Gang:2024abc}. We think it will be an interesting research topic to figure out what UV line operators correspond to simple objects of the $\TFTk$ for the case when $p$ is even and using them to construct full modular matrices.

Another issue associated with the $\TFTk$ is a mismatch between $(t,\vec{n}_i)$ labels of Bethe vacua and simple objects for the case when $p$ is odd. See appendix~\ref{Ap: Explicit TFTk}. If one can clarify how the $\TFTk$ is embedded in the $D(\vec{k})|_A$ theory, it will be possible to introduce natural labeling for both of them which makes
\begin{align}
    T'(t',i')=t'\;,\quad \vec{n}'(t',i')=\vec{n}_{i'}\;.
\end{align}

\paragraph{Refined ansatz for problematic cases} Since Bethe vacua obtained from the ansatz \eqref{eq: Ansatz} are inconsistent with the assumption in some cases, we have to exclude them. We are exploring the possibility of designing an alternative double-scaling ansatz, with the expectation that it will allow us to deal with all excluded cases. See \cite{Jeong:2026dzz} for a recent study in this direction.

\acknowledgments
We are grateful to Dongmin Gang for his early collaboration and also for helpful discussions and comments at various stages of the project. We also thank Heesu Kang and Sungjoon Kim for helpful discussions. The work of KJ and SL is supported in part by the National Research Foundation of Korea grant NRF-2022R1C1C1011979. We also acknowledge the support by the National Research Foundation of Korea grant RS-2024-00405629.

For this version, we would like to thank Huijoon Sohn for many helpful discussions during the preparation of the next project.

\appendix
\section{WRT TQFT} \label{Ap: WRT review}
The Witten-Reshetikhin-Turaev(WRT) invariants are topological invariants of 3-manifolds arising from the ${\rm SU}(2)$ Chern-Simons theory. Witten showed that the path integral quantization of the ${\rm SU}(2)$ Chern-Simons theory at level $k$ yields 3-manifold invariants and knot polynomials, with modular $S$ and $T$ matrices computed from the ${\rm SU}(2)_k$ Wess-Zumino-Witten(WZW) characters\cite{Witten:1988hf}. Moore and Seiberg identified the underlying structure as the Modular Tensor Category(MTC) which is characterized by its simple objects, braiding and fusion rules\cite{Moore:1988uz}.
On the other hand, Reshetikhin and Turaev provided an algebraic construction of MTC from the quantum group \(U_x\mathfrak{sl}(2,\mathbb{C})\) at primitive root of unity \(x\) over field \(\mathbb{C}\) and the whole process was further systemized by Turaev\cite{Reshetikhin:1990tk,Reshetikhin:1991tc,Turaev:1994nc}. It is widely noticed that the special case when \(x=e^{\frac{2\pi i}{k+2}}\) is equivalent to \({\rm SU}(2)_k\) Chern-Simons theory. However, it was the first (as we know) mathematically rigorous realization of non-unitary TQFTs by choosing other primitive roots of unity.

\subsection{Notes on the RT construction}\label{Subsec: RT construction}
Reshetikhin-Turaev construction is a sequence of stacking structures to result in the MTC and further to the TQFT. The skeleton of the procedure is as follows. 
\begin{center}
    \begin{tikzpicture}[label/.style={above, font=\scriptsize}]
        \node (0){\(D(U_x\mathfrak{b}_+)\)};
        \node (1)[right=0.5cm of 0]{\(U_x\mathfrak{g}\)};
        \node (2)[right=1.5cm of 1]{\(\mathcal{C}\supset \mathcal{T}\)};
        \node (3)[right=1cm of 2]{\(\mathrm{MTC}\)};
        \node (4)[right=3cm of 3]{\(\mathrm{TQFT}=:\mathrm{RT}(x,\mathfrak{g})\)};    

        \draw (1.north)edge[->,out=135, in=45, looseness=1]node[label]{Drinfeld Double}(0.north);
        \draw (0.south)edge[->,out=315, in=225, looseness=1]node[label,below]{\(\cdot/q\textrm{-Serre eq.}\)}(1.south);
        \draw[->] (1.east)--node[label]{\(\textrm{Rep}^{\textrm{f.d.}}(\cdot)\)}(2.west);
        \draw[->] (2.east)--node[label]{\(\cdot/\mathcal{N}\)}(3.west);
        \draw[->] (3.east)--node[label]{Reshetikhin-Turaev}(4.west);
    \end{tikzpicture}
\end{center}
Essentially, the MTC is defined to be a finite, non-degenerate and semi-simple braided category. As noted in the diagram, we denote the resulting TQFT from the quantum group of \(\mathfrak{g}\) at root of unity \(x\) as \(\textrm{RT}(x,\mathfrak{g})\).

The braided structure is granted by realizing the quantum group \(U_x\mathfrak{g}\) as a quasi-triangular Hopf algebra. Such a procedure is known as the ``Drinfeld double trick"\cite{Drinfeld:1988jsm}. In specific, one defines the coassociative comultiplication \(\Delta\) to find a universal invertible \(\mathcal{R}\) matrix that satisfies some constraint equations composed with two the hexagon axioms and one for compatibility with \(\Delta\). The coassociativity of \(\Delta\) naturally induces the pentagon axiom. 

Finiteness and semi-simplicity are saturated during the quotient by \(\mathcal{N}\), the category of the \textit{negligible morphisms} between \(V,W\in\textrm{Obj}(\mathcal{C})\) defined as
\begin{align}
    \mathcal{N}(V,W)=\{f\in \textrm{Hom}(V,W) \,\vert\,\textrm{Tr}_x(g\circ f) =0\;, {}^{\forall}g\in\textrm{Hom}(W,V)\}\;.
\end{align}
At roots of unity, the category of all finite dimensional representations is non-semi-simple. A key feature of the problematic representations in $\mathcal{C}$ which obstruct us from proceeding to the next step is that their quantum dimension \([\cdot]_x\) is zero. To cure this, one restricts attention to the subcategory of tilting modules \(\mathcal{T}\), a class of well-behaved representations that remains closed under tensor products. The MTC is then obtained by taking the quotient\footnote{This procedure generally yields the maximal semi-simple category. That is, one can further quotient by some object to obtain smaller semi-simple category.} of \(\mathcal{T}\) by $\mathcal{N}$.

It is important to note that the defining data of MTC, the \(R\), \(F\) matrices and the twist \(\theta\) are derived from \(\Delta\) and \(\mathcal{R}\). Specifically, for chosen proper basis, the matrix representation of \(\mathcal{R}\) corresponds to the \(R\) matrix and the eigenvalue of it corresponds to the \(\theta\). The \(F\) matrix arises from the associator, representing the base change between different fusion orders allowed by the coassociativity of \(\Delta\). Explicitly, the pentagon and hexagon equations are given by 
\begin{align}\begin{split} \label{eq:penthex}
    [F_{e}^{fcd}]_{g\ell} [F_{d}^{ab\ell}]_{fk} &= \sum_{h} [F_{g}^{abc}]_{fh} [F_{e}^{ahd}]_{gk} [F_{k}^{bcd}]_{h\ell}\;,\\
    (R_{e}^{ac})^{\pm 1} [F_{d}^{acb}]_{eg} (R_{g}^{bc})^{\pm 1} &= \sum_{f} [F_{d}^{cab}]_{ef} (R_{d}^{fc})^{\pm 1} [F_{d}^{abc}]_{fg}
\end{split}\end{align}
respectively\cite{Aboumrad2022:qc}. Note that equations are in polynomial form. It will play an important role in Galois conjugation introduced in following appendix. 

\subsection{Galois Conjugation : Algebraic vs. Categorical Levels}
In RT construction, the fact that \(x\) is a root of unity plays a significant role in making the finite semi-simple category. However, there is no further restriction on which root of unity should be selected and in fact, any root of unity gives a proper MTC. Hence, one can consider a Galois conjugation on \(x\) to obtain a similar but different MTC from the original one, usually one from the principal root of unity. This Galois conjugation can be done before the RT construction. The Galois group is given by
\begin{equation}
    \textrm{Gal}_A:=\textrm{Gal}(\mathbb{Q}(\zeta_r)/\mathbb{Q})\cong (\mathbb{Z}/r\mathbb{Z})^\times
\end{equation}
where \(\zeta_r\) is a principal \(r^{\textrm{th}}\) root of unity. Elements in such a group act as 
\begin{equation}
    \sigma_m (\textrm{RT}(x,\mathfrak{g})) :=\textrm{RT}(\sigma_m (x), \mathfrak{g}) = \textrm{RT}(x^m,\mathfrak{g});, \quad \sigma_m \in \textrm{Gal}_A\;.
\end{equation}
Let's call it an algebraic level Galois conjugation.

On the other hand, recall that the pentagon and hexagon axioms \eqref{eq:penthex} appear as a $\mathbb{Q}$-coefficient multi-variable polynomial equation on a proper basis. Furthermore, even though there is a gauge degree of freedom associated with choosing different basis in fusion space, it has been shown that there is a gauge choice such that the modular data \(R\), \(F\), and even \(S\) and \(T\) matrices lie on the same finite number field\cite{Davidovich:2013amc}. Therefore, following an argument of \cite{Buican:2023pme}, we can consider categorical level Galois conjugation through an element of the following Galois group
\begin{equation}
\textrm{Gal}_C:=\textrm{Gal}(\mathbb{Q}(\zeta_{r'})/\mathbb{Q})\cong (\mathbb{Z}/r'\mathbb{Z})^\times
\end{equation}
with \(r' \in r\mathbb{Z}\) such that \(\mathbb{Q}(\zeta_{r'})=\mathbb{Q}(S,T)\). Elements in such a group act as
\begin{equation}
    \sigma_m (\textrm{RT}(x,\mathfrak{g})) := (\sigma_m (S),\sigma_m (T), \cdots)\;, \quad \sigma_m \in \textrm{Gal}_C
\end{equation}
when identifying TQFT theory as a tuple of defining data such as modular \(S\) and \(T\) matrices.

It is important to note that there are MTCs that can only be derived from the algebraic level Galois conjugation and that from the categorical level Galois conjugation. The former one is clear as \(r'=\# r\) enhances the condition  of \(\sigma_m\) from \(\textrm{gcd}(m,r)=1\) to \(\textrm{gcd}(m,r')=1\). There are in fewer \(m\) that is coprime to \(r'\) than \(r\) in \(\{1,2,\cdots,r\}\). The latter one is from the fact that \(r'\) is larger than \(r\). There are possible \(m'=m+ nr\) for \(n\in\mathbb{Z}\) such that \(0\leq n<\#\). MTCs from such Galois conjugation are not obtainable from choosing a proper root of unity at the beginning of RT construction. 

\subsection{\(\textrm{RT}(x,\mathfrak{sl}(2,\mathbb{C}))\)}
In this paper, we focus only on the case \(\mathfrak{g}=\mathfrak{sl}(2,\mathbb{C})\). Recall that \(\textrm{RT}(\zeta_{k+2},\mathfrak{sl}(2,\mathbb{C}))\) is the \({\rm SU}(2)_k\) Chern-Simons theory whose simple objects correspond to the irreducible representations of \(\mathrm{SU}(2)_k\), labeled by spin \(j \in \{0, \frac{1}{2}, 1, \dots, \frac{k}{2}\}\). These objects are denoted as \(\ket{j}\). Modular \(S\) and \(T\) matrices are given by
\begin{align}\label{eq: Original wrt S and T}
\frac{S_{\alpha\beta}}{S_{11}}=\frac{x^{\alpha\beta/2}-x^{-\alpha\beta/2}}{x^{1/2}-x^{-1/2}}\;,\quad T_{\alpha\beta}= \delta_{\alpha\beta}x^{\alpha^2/4}
\end{align}
for \(\alpha,\beta\in\{1,2,\dots,k+1\}\). From the definition of $T_{\alpha\beta}$, we can find that the number field in which the modular data live is \(\mathbb{Q}(\zeta_{4(k+2)})\), that is, \(r'=4r\). As we discussed earlier, we can choose any \(m\) that is coprime to \(k+2\) from \(1\) to \(4(k+2)\) to make a Galois conjugation \(\sigma_m\) and in general, they will transform the modular data by just replacing \(x\) to \(x^m\). But we need to be careful for several non-trivial cases. We will denote the conjugated TQFT as \(\textrm{WRT}_k^m\). 

\subsubsection{Odd \(k\) : integer-spin subcategory}
A particularly interesting structure emerges when the level \(k\) is an odd integer. In this case, the set of simple objects \(\{\ket{j}\}_{j=0}^{k/2}\) naturally splits into two sectors: One with integer spins \(j \in \{0, 1, \dots, (k-1)/2\}\), and the other one with half-integer spins \(j \in \{1/2, 3/2, \dots, k/2\}\). The fusion rules exhibit a \(\mathbb{Z}_2\)-grading with respect to this decomposition. The key object responsible for this symmetry is the highest-spin representation \(\ket{k/2}\). Because \(k\) is odd, \(k/2\) is a half-integer. This simple object is an \emph{invertible line}, or a simple current, meaning that its fusion inverse exists. Specifically, its fusion square is the tensor unit
\begin{align}
    \ket{k/2} \otimes \ket{k/2} \cong \ket{0}\;.
\end{align}
Tensoring any object with this current \(\ket{k/2}\) implements the \(\mathbb{Z}_2\) symmetry, mapping integer-spin objects to half-integer spin objects (\(\ket{j} \otimes \ket{k/2}=\ket{k/2-j}\)) and vice-versa.

We can then perform a procedure known as `gauging' or `orbifolding' this \(\mathbb{Z}_2\) symmetry. This amounts to restricting the category to the objects in the trivial sector of the grading, which are the integer-spin representations. This subcategory is itself a consistent, unitary MTC.

\subsubsection{Even \(m\) : non-trivial Galois conjugation}
Recall that the categorical level Galois conjugation of MTC is associated with \(\mathrm{Gal}_C=\textrm{Gal}(\mathbb{Q}(\zeta_{4(k+2)})/\mathbb{Q})\) which does not admit an even \(m\) since it requires \(\textrm{gcd}(m,4(k+2))=1\). However, the algebraic level Galois conjugation allows \(\textrm{Gal}(\mathbb{Q}(\zeta_{k+2})/\mathbb{Q})\) transformed theory. This exactly corresponds to the case when \(m\) is even and therefore \(k\) is odd. This choice of \(m\) has non-trivial consequences for the properties of the MTC. The quantum dimension of a simple object \(\ket{j}\) is given by
\begin{align}
    \dim_x(\ket{j}) = [2j+1]_x = \frac{\sin\bigr(\frac{\pi m(2j+1)}{k+2}\bigr)}{\sin\bigr(\frac{\pi m}{k+2}\bigr)}\;.
\end{align} 
There is the crucial identity that connects quantum dimensions of two objects whose spins are symmetric with respect to \(k/4\)
\begin{align}
     \dim_x(\ket{k/2-j}) = (-1)^{m-1} \dim_x(\ket{j})\;.
\end{align}
This relation governs the modular properties of the theory. When \(m\) is odd, \(m-1\) is even, and the relation becomes symmetric:
\begin{align}
    \dim_x(\ket{k/2-j}) = \dim_x(\ket{j})\;.
\end{align}
The \(S\) matrix is non-degenerate, and both the full \(\mathrm{SU}(2)_k\) category and its integer-spin subcategory (for odd \(k\)) are well-defined. The situation is drastically different when \(m\) is even. The condition \(\text{gcd}(m, k+2)=1\) implies that \(k\) must be odd. For even \(m\), the identity becomes anti-symmetric:
\begin{align}
\dim_x(\ket{k/2-j}) = - \dim_x(\ket{j})\;.
\end{align}
This is not merely a feature, but a direct cause of degeneracy. This anti-symmetricity reflects a linear dependence between the columns of the \(S\) matrix itself, forcing its determinant to be zero. The presence of these \textit{ghost} objects, where the direct sum \(\ket{j} \oplus \ket{k/2-j}\) has zero total quantum dimension, enlarges the negligible category \(\mathcal{N}\). Fortunately, since \(k\) is odd, \(j\) is an integer if and only if \(k/2-j\) is a half-integer. The pairing of objects with opposite-sign quantum dimensions always occurs between the integer-spin and half-integer spin sectors. Therefore, an additional quotient by that enlarged part can be understood as orbifolding the \(\mathbb{Z}_2\) symmetry, restricting only to the integer-spin subcategory.

\subsubsection{Modular \(S\) and \(T\) matrices}
\label{Subsubsec: WRT S and T}
The equivalence of two MTCs is strongly indicated by identical modular $S$ and $T$ matrices. These matrices are often more accessible from a physics perspective than the full MTC data, such as $R$ and $F$ matrices. This appendix summarizes \(S\) and \(T\) matrices for \(\textrm{WRT}_k^m\) and its relevant variations. Here, we introduce the sign factor $\epsilon$ defined as
\begin{align} \label{eq: Sign factor for WRT S}
    \epsilon\equiv(-1)^{N+(m+1)(m'+1)}\quad\textrm{with}\quad mm'=(k+2)N+1, \quad m'\in\{1,2,\cdots, k+1\}
\end{align}
for given $m\in\{1,2,\cdots,k+1\}$ to ensure that at least one row of the $S$ matrix becomes positive, from the physical motivation\cite{Gannon:2003kg}. The phase factor $\xi$ can be determined up to $e^{\frac{2\pi i}{3}\mathbb{Z}}$, from the ${\rm SL}(2,\mathbb{Z})$ relation $S^2=(ST)^3$.
\paragraph{Odd $m$} For $\a,\b\in\{1,2,\cdots,k+1\}$,
\begin{align}
    S_{\a\b}=\e\sqrt{\frac{2}{k+2}}\sin\bigr(\frac{m\pi\a\b}{k+2}\bigr)\;,\quad T_{\a\b}=\xi\d_{\a\b}\exp\bigr(\frac{\pi im\a^2}{2(k+2)}\bigr)\;.
\end{align}
We also have
\begin{itemize}
    \item $2(k+2)+m$
    \begin{align}
    S_{\a\b}=\e\sqrt{\frac{2}{k+2}}\sin\bigr(\frac{m\pi\a\b}{k+2}\bigr)\;,\quad T_{\a\b}=\xi\d_{\a\b}\exp\bigr(\frac{\pi im\a^2}{2(k+2)}+i\pi \a\bigr)\;.
    \end{align}
    \item $4(k+2)-m$
    \begin{align}
    S_{\a\b}=\e\sqrt{\frac{2}{k+2}}\sin\bigr(\frac{m\pi\a\b}{k+2}\bigr)\;,\quad T_{\a\b}=\xi\d_{\a\b}\exp\bigr(-\frac{\pi im\a^2}{2(k+2)}\bigr)\;.
    \end{align}
    \item $2(k+2)-m$
    \begin{align}
    S_{\a\b}=\e\sqrt{\frac{2}{k+2}}\sin\bigr(\frac{m\pi\a\b}{k+2}\bigr)\;,\quad T_{\a\b}=\xi\d_{\a\b}\exp\bigr(-\frac{\pi im\a^2}{2(k+2)}-i\pi \a\bigr)\;.
\end{align}
\end{itemize}
\paragraph{Even $m$} In this case, $k$ should be \textit{odd} and as we discussed, WRT construction requires a further quotient into integer-spin subcategory. For $a,b\in\bigr\{1,2,\cdots,\frac{k+1}{2}\bigr\}$, we have
\begin{align}
    \bar{S}_{ab}=\e\sqrt{\frac{4}{k+2}}\sin\bigr(\frac{m\pi(2a-1)(2b-1)}{k+2}\bigr)\;,\quad \bar{T}_{ab}=\xi\d_{ab}\exp\bigr(\frac{2\pi ima(a-1)}{k+2}\bigr)\;.
\end{align}
We also have $2(k+2)-m$,
\begin{align}
    \bar{S}_{ab}=\e\sqrt{\frac{4}{k+2}}\sin\bigr(\frac{m\pi(2a-1)(2b-1)}{k+2}\bigr)\;,\quad \bar{T}_{ab}=\xi\d_{ab}\exp\bigr(-\frac{2\pi ima(a-1)}{k+2}\bigr)\;.
    \end{align}

\section{Essential methods}
\label{Ap: Methods}
\subsection{Topological twist and Bethe/Gauge data} \label{Ap: Topl twist}
Consider the 3d $\CN=4$ theory $\CT$. It has ${\rm SO}(\CN=4)_{R}\cong {\rm SU}(2)_C\times {\rm SU}(2)_H$ R-symmetry, where the subscripts $C$ and $H$ indicate \textit{Coulomb} and \textit{Higgs}, respectively. There are two branches of the moduli space relevant to our discussion. One is the Coulomb branch where chiral primary operators are charged under only ${\rm SU}(2)_C$ and the other one is the Higgs branch, charged under only ${\rm SU}(2)_H$\cite{Gang:2021hrd}. In terms of $\mathcal{N}=2$ subalgebra, two Cartan generators $J^C_3$ and $J^H_3$ become generators of $\mathrm{U}(1)_R$ R-symmetry and $\mathrm{U}(1)_A$ flavor symmetry through combinations
\begin{align}
    \mathrm{U}(1)_R:J_3^C+J_3^H\;,\quad \mathrm{U}(1)_A:J^C_3-J^H_3\;.
\end{align}
Here, $\mathrm{U}(1)_R$ indicates the superconformal R-symmetry in the IR\cite{Gang:2021hrd,Jafferis:2012iv}. We define the mixing parameter $\nu\in\mathbb{R}$ between ${\rm U}(1)_R$ and ${\rm U}(1)_A$ as
\begin{align}
    J^R_\nu\equiv J_3^C+J_3^H+\nu(J_3^C-J_3^H)
\end{align}
to consider a different choice of the R-symmetry.

Assuming $\CT$ is a rank 0 theory which has an empty Coulomb and Higgs branches\cite{Gang:2018huc}, let's assign a different R-symmetry corresponding to $\nu=-1$, and also set the real mass parameter $M$ of the vector multiplet associated with ${\rm U}(1)_A$ as zero. This procedure is called (topological) A-twist and the resulting theory $\CT|_A$ becomes a non-unitary TQFT\cite{Gang:2024wxz,Gang:2021hrd}.

To extract the TQFT data of $\CT|_A$, it has proven to be useful to consider the supersymmetric localization\cite{Pestun:2016zxk} on squashed 3-sphere backgrounds labeled by the squashing parameter $b^2>0$:
\begin{align}\label{eq: Squashed 3-sphere}
    S^3_b\equiv\biggr\{x\in\mathbb{R}^4\;\biggr|\;b^2\bigr((x^1)^2+(x^2)^2\bigr)+\frac{1}{b^2}\bigr((x^3)^2+(x^4)^2\bigr)=1\biggr\}\;.
\end{align}
It allows us to find a set of Bethe vacua ${\rm BV}^{\CT}$ and the $\HF$ data through the Bethe/Gauge analysis as shown in section~\ref{Subsec: D(k) BG analysis}. Then $\HF^{\CT}$ is identified with partial modular data of TQFT $\CT|_A$\cite{Gang:2021hrd} as
\begin{align}
    \HF^{\CT}=\bigr\{(\CH,\CF)\bigr\}=\bigr\{(|S_{1\a}|^{-2},\xi T_{\a\a})\bigr\}\;,\quad \xi\in e^{i\mathbb{R}}
\end{align}
where $S$ and $T$ are modular matrices of $\CT|_A$. A set of Handle-gluings is identified with the first row(=first column) of the modular $S$ matrix
\begin{align}
    \bigr\{|\CH|^{-\frac{1}{2}}\bigr\}=\bigr\{|S_{1\a}|\bigr\}\;.
\end{align}
Since $|S_{11}|$ encodes the 3-sphere partition function of the TQFT, we also find
\begin{align} \label{eq: S11 from HF data}
    |S_{11}|=\bigr|Z^{S^3}_{\CT|_A}\bigr|=\biggr|\sum_{\HF}\CH^{-1}\CF\biggr|=\bigr|\CH(V_1)\bigr|^{-\frac{1}{2}}
\end{align}
where $V_1$ indicates the true Bethe vacuum. See around \eqref{eq: True vac condition} for its definition. Note that \eqref{eq: S11 from HF data} implies the characteristic property of $\HF$ data \eqref{eq: HF data relation}
\begin{align}
    \bigr(\bigr|Z^{S^3_b}\bigr|=\bigr)\biggr|\sum_{\HF}\CH^{-1}\CF\biggr|=\bigr|\CH_1\bigr|^{-\frac{1}{2}}\text{ for some }\CH_1\in\HF
\end{align}
we used in section~\ref{Sec: WRT data extraction}.

We can also compute the \textit{index} associated with the Hilbert space constructed on the Riemann surface of genus $g$ : $H(\S_g)$ using $\HF$ data as\cite{Benini:2015noa,Closset:2016arn}
\begin{align} \label{eq: TTI}
    \CI(\S_g)[\mathcal{X}]\equiv {\rm Tr}_{H(\S_g)}\bigr((-1)^R\mathcal{X}\bigr)=\sum_{V\in\frac{{\rm BV}^{\CT}}{(\textbf{Weyl})}}\CH^{g-1}(V)\mathcal{X}(V)\;.
\end{align}
Here, $\mathcal{X}$ indicates the inserted (gauge invariant) observable preserving the Supersymmetry.

\subsection{Determination of the full modular matrix} \label{Ap: Method full modular matrix}
A TQFT has a special set of framed line observables called \textit{simple object}(\textit{line}). In this paper, the terms ``simple object" of a TQFT and associated MTC will be used interchangeably. This is because the two are connected by the dictionary\cite{Witten:1988hf}
\begin{align}
\label{eq: S matrix from line}
    \int[D\Psi]e^{-S^E_{\mathrm{TQFT}}[\Psi;S^3]}W^\alpha_{\mathcal{C}_1}W^{\beta}_{\mathcal{C}_2}=S^{\mathrm{TQFT}}_{\alpha\beta}
\end{align}
where $W^\alpha_{\CC}$ indicates the zero-framing simple line following the curve $\mathcal{C}$ in $S^3$ and the two curves $\mathcal{C}_{1,2}$ form the Hopf link configuration as shown in figure~\ref{fig:main problem} in the introduction.

For the TQFT $\mathcal{T}|_A$ constructed through the topological twist, we have Wilson line observables at the UV labeled by its Gauge/Flavor representations. Even though there are infinitely many representations, the IR appropriate equivalence relations are induced and as a result some of them form simple objects for the TQFT\cite{Gang:2021hrd}. RG invariant computations allow us to test the equivalence relations explicitly. For example, the superconformal index\cite{Kim:2009wb} provides us strong criteria\cite{Gang:2024abc}
\begin{align} \label{eq: Simple crit sci}
    \begin{split}
        &[\text{Criterion 1}]
        \\&W^{\a}\neq 1({\rm Triv.})\text{ becomes the simple object in the IR}
        \\&\Longrightarrow \lg W^{\a}_{N,S}\rg_{\rm Sci.}(q,\eta=1,\n=-1)=0\;,\quad \lg W^{\a}_NW^{\a}_{S}\rg_{\rm Sci.}(q,\eta=1,\nu=-1)=1\;.
        \\
        \\&[\text{Criterion 2(Overlap test)}]
        \\&W^\a\text{ and }W^{\b\neq \a}\text{ are independent objects in the IR}
        \\&\Longrightarrow \lg W^{\a}_N W^{\b}_S\rg_{\rm Sci.}(q,\eta=1,\n=-1)=0\;.
    \end{split}
\end{align}
Here, $\eta$ and $\nu$ indicate the fugacity associated with $\mathrm{U}(1)_A$ and the mixing parameter between ${\rm U}(1)_R$ and ${\rm U}(1)_A$. $N$ and $S$ indicate two possible curves allowed in the localization on the Superconformal index background $S^2\times S^1$\cite{Kapustin:2009kz,Closset:2018ghr}. Alternatively, we use the topological twisted index \eqref{eq: TTI} when $g=0$, which provides the $q=1$ version of \eqref{eq: Simple crit sci}. It is better suited to our setting since we have $\HF$ data. 
\begin{align} \label{eq: Simple crit TTI}
    \begin{split}
        &[\text{Criterion 1}]
        \\&W^{\a}\neq 1({\rm Triv.})\text{ becomes the simple object in the IR}
        \\&\Longrightarrow \sum_{V\in\frac{{\rm BV}^\CT}{(\textbf{Weyl})}}\CH^{-1}(V)W^\a_{\CC_+}(V)=0\;,\quad \sum_{V\in\frac{{\rm BV}^\CT}{(\textbf{Weyl})}} \CH^{-1}(V)W^\a_{\CC_+}(V)\bigr(W^\a_{\CC_+}(V)\bigr)^*=1\;.
        \\
        \\&[\text{Criterion 2(Overlap test)}]
        \\&W^\a\text{ and }W^{\b\neq \a}\text{ are independent objects in the IR}
        \\&\Longrightarrow \sum_{V\in\frac{{\rm BV}^\CT}{(\textbf{Weyl})}}\CH^{-1}(V) W^\a_{\CC_+}(V)\bigr(W^\b_{\CC_+}(V)\bigr)^*=0\;.
    \end{split}
\end{align}
Here, $\mathcal{C}_+$ indicates one of the two possible curves allowed in the localization on the squashed 3-sphere background which its length is $2\pi b$\cite{Kapustin:2009kz,Closset:2018ghr}.

\begin{figure}[t]
\centering
\includegraphics[width=1\textwidth]{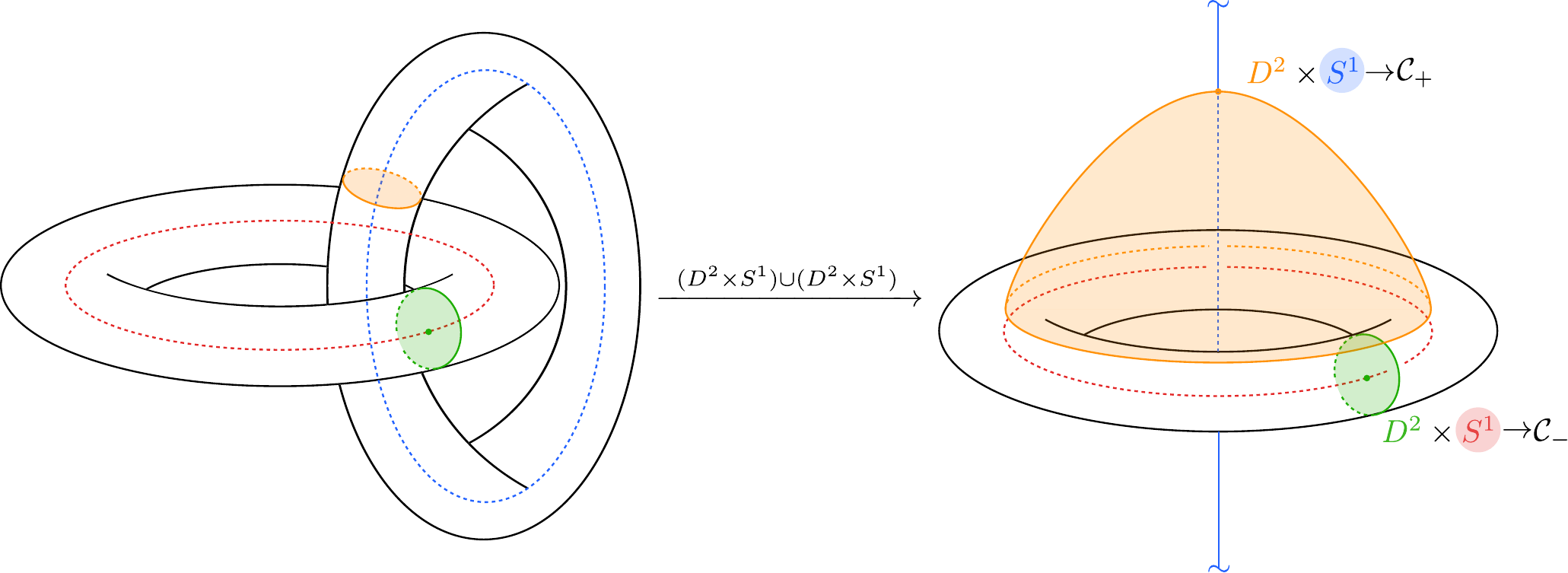}
\captionof{figure}{Cartoon of two curves $\mathcal{C}_{\pm}\in S^3_b$. Note that $S^3\overset{\mathrm{Topo.}}{\cong} (D^2\times S^1)\cup (D^2\times S^1)$.}
\label{fig: SUSY lines in S3b}
\end{figure}
Criteria in \eqref{eq: Simple crit TTI} provide only sufficient conditions for simple objects. However, suppose that we somehow have a correct \textit{maximally independent set} of extended observables, especially Wilson lines that flow to the IR TQFT's simple objects. Here, a maximally independent set means a set of extended observables whose number is equal to that of the Bethe vacua and satisfy criteria \eqref{eq: Simple crit TTI}. Then we can use the dictionary \eqref{eq: S matrix dict} derived in~\cite{Gang:2021hrd} to determine the modular $S$ matrix. Accepting the conjecture introduced in~\cite{Gang:2025}, we can understand it as follows. Let's try to determine the modular $S$ matrix through the dictionary \eqref{eq: S matrix from line} as
\begin{align}
\begin{split}
    Z^{S^3_b}_{\mathcal{T}|_A}[W^\alpha_{\mathcal{C}_-}W^{\beta}_{\mathcal{C}_+}]\xrightarrow[b^2\rightarrow0]{\text{SUSY Localization}}&\int d\vec{Y}e^{\frac{1}{\hbar}\mathcal{W}_0^{\mathcal{T}|_A}(\vec{Y})+\cdots}W^\alpha_{\mathcal{C}_-}W^{\beta}_{\mathcal{C}_+}
    \\&\equiv\int d\vec{Y}\exp\biggr(\frac{\tilde{\mathcal{W}}_0^{\mathcal{T}|_A}(\vec{Y})}{\hbar}+\tilde{\mathcal{W}}_1^{\mathcal{T}|_A}(\vec{Y})+\mathcal{O}(\hbar)\biggr)
    \\&\propto S_{\alpha\beta}
\end{split}
\end{align}
where $W^{\alpha=1}\equiv 1({\rm Triv.})$. Here, $\mathcal{C}_-$ indicates the other curve allowed in the localization on the squashed 3-sphere background, which its length is $2\pi/b$ and forms the Hopf link configuration with $\mathcal{C}_+$. See figure~\ref{fig: SUSY lines in S3b}. Two Wilson lines play a significantly different role due to their length difference\cite{Kapustin:2009kz,Closset:2018ghr}.
\begin{itemize}
    \item $\mathcal{W}^\alpha_{\mathcal{C}_-}$ contributes to $\tilde{\mathcal{W}}^{\mathcal{T}|_A}_0$ and shifts the true Bethe vacuum from $V_1$ to $V_\alpha$\cite{Gang:2025}. It defines the Bethe vacuum-Simple object correspondence
    \begin{align}
        \label{eq: BVSO map}
        V_1\longleftrightarrow 1(\mathrm{Triv.})\;,\quad V_{\alpha}\longleftrightarrow  W^\alpha\;.
    \end{align}
    \item $\mathcal{W}^\beta_{\mathcal{C}_+}$ contributes to $\tilde{\mathcal{W}}_1^{\mathcal{T}|_A}$.
\end{itemize}
So the modular $S$ matrix can be computed as
\begin{align} \label{eq: S matrix condition starting point}
    S_{\alpha\beta}=\zeta_{\a} W^{\b}_{\CC_+}(V_\a)|\CH(V_\a)|^{-\frac{1}{2}}
\end{align}
where $\zeta_\alpha$ is the phase factor associated with the choice of the true Bethe vacuum. It is generally difficult to determine the Bethe vacuum $V_\alpha$ and $\zeta_\alpha$ corresponding to a given loop operator $W^\alpha_{\mathcal{C}_-}$. However, correspondence with a boundary CFT in mind, we can detour this issue by using some physical properties of the modular $S$ matrix.
\begin{itemize}
    \item The modular $S$ matrix should be a symmetric matrix\cite{Wen:2015qwa,Gannon:2003kg}. Especially, the condition for the first row(=first column)
    \begin{align}
        S_{1\a}=S_{\a 1}\longrightarrow \zeta_1 W^\a_{\CC_+}(V_1)|\CH(V_1)|^{-\frac{1}{2}}=\zeta_\alpha|\CH(V_\alpha)|^{-\frac{1}{2}}
    \end{align}
    provides us the useful relation
    \begin{align} \label{eq: S matrix dict}
        S_{\a\b}=W^\b_{\CC_+}(V_\a)S_{1\a}\;.
    \end{align}
    \item At least one row of the modular $S$ matrix should be a positive due to physical requirements associated with the \textit{minimal primary} appearing in non-unitary CFT\cite{Gannon:2003kg}. For the unitary theory which satisfies $|S_{11}|\le|S_{1\alpha}|$\cite{Gang:2021hrd}, it must be the first row due to its quantum dimension interpretation\cite{Wen:2015qwa}. 
    \item The fusion coefficient which describes the relation
    \begin{align}
        \mathcal{O}_\alpha\otimes \mathcal{O}_\beta=\bigoplus_{\gamma}N^\gamma_{\alpha\beta}\mathcal{O}_\gamma\;,\quad N^\gamma_{\alpha\beta}=\sum_{\omega}\frac{S_{\alpha\omega}S_{\beta\omega}(S_{\gamma\omega})^*}{S_{1\omega}}
    \end{align}
    should satisfy $N^\gamma_{\alpha\beta}\in\mathbb{N}_0$\cite{Wen:2015qwa,Verlinde:1988sn}.
\end{itemize}
With the definition of the true Bethe vacuum $V_1$, they almost uniquely determine the modular $S$ matrix. Combine it with the other dictionary for the modular $T$ matrix\cite{Gang:2021hrd}
\begin{equation}
\label{eq: T matrix dict}
    T_{\alpha\beta}=\xi \delta_{\alpha\beta}\mathcal{F}(V_\alpha)\;,
\end{equation}
we can determine the full modular matrices of the TQFT $\mathcal{T}|_A$. Note that with the proper choice of the phase factor $\xi$, $S$ and $T$ should satisfy the ${\rm SL}(2,\mathbb{Z})$ relation\cite{Wen:2015qwa,Gannon:2003kg}
\begin{equation}
\label{eq: SL(2,Z) condition}
    S^2=(ST)^3= C\;,\quad C^2=1\;,\quad C_{\alpha\beta}\equiv N^1_{\alpha\beta}\;.
\end{equation}

\section{Detail of the Bethe/Gauge analysis}
\label{Ap: BG detail}
\subsection{$T[\mathrm{SU}(2)]$ theory}
\paragraph{Squashed 3-sphere partition function} At the superconformal point $\nu=0$, the squashed 3-sphere partition function for the $T[\mathrm{SU}(2)]$ theory is evaluated as ($\hbar\equiv 2\pi ib^2$) 
\begin{align}
\label{eq: tsu2 partition function}
    \begin{split}
        &Z_{T[\mathrm{SU}(2)]}^{S_b^3}[W_1,W_2;M,\nu=0]=\xi \int \frac{dZ}{\sqrt{2\pi\hbar}} e^{\frac{Z^2 + 2ZW_2 + W_1^2}{\hbar}}\psi_\hbar\left[-M + \left(\pi i + \frac{\hbar}{2}\right)\right]\\
        &\qquad\quad\psi_\hbar\left[Z+W_1+\frac{1}{2}M + \frac{1}{2}\left(\pi i + \frac{\hbar}{2}\right)\right]\psi_\hbar\left[Z-W_1+\frac{1}{2}M + \frac{1}{2}\left(\pi i + \frac{\hbar}{2}\right)\right]\\
        &\qquad\quad\psi_\hbar\left[-Z+W_1+\frac{1}{2}M + \frac{1}{2}\left(\pi i + \frac{\hbar}{2}\right)\right]\psi_\hbar\left[-Z-W_1+\frac{1}{2}M + \frac{1}{2}\left(\pi i + \frac{\hbar}{2}\right)\right]\;.
    \end{split}
\end{align}
See appendix~\ref{Ap: QDL} for the definition of the quantum dilogarithm function $\psi_\hbar (Z)$. $\xi$ is the subtle phase factor. $W_1$ and $W_2$ are ($2\pi b$-multiplied) real mass parameters for ${\rm SU}(2)_L$ and ${\rm SU}(2)_R$ respectively. $M$ is the rescaled real mass parameter for ${\rm U}(1)_A$. We can assign different R-charges for chirals at the UV as follows :
\begin{align}
    Z^{S^3_b}_{T[\mathrm{SU}(2)]}[W_1,W_2;M,\nu]=Z^{S^3_b}_{T[\mathrm{SU}(2)]}[W_1,W_2;M+\left(\pi i+\frac{\hbar}{2}\right)\nu,0]\;.
\end{align}
From now on, we are going to set \(M=0\) and simply denote
\begin{align}
    Z^{S^3_b}_{T[\mathrm{SU}(2)]}[W_1,W_2,;M=0,\nu]\longrightarrow Z_{T[\mathrm{SU}(2)]}[W_1,W_2;\nu]\;.
\end{align}

\paragraph{$b^2\rightarrow 0+$ expansion} Using
\begin{align}
    \frac{d}{dX}\tilde{\mathrm{Li}}_2(e^{-X})=-\tilde{\mathrm{Li}}_1(e^{-X}) \rightarrow\tilde{\mathrm{Li}_2}(e^{-X+\hbar Y})=\tilde{\mathrm{Li}}_2(e^{-X})+\hbar Y \tilde{\mathrm{Li}_1}(e^{-X})+\mathcal{O}(\hbar ^2)\;,
\end{align}
we can expand $Z_{T[\mathrm{SU}(2)]}[W_1,W_2;\nu]$ as
\begin{align}
\label{eq: tsu2 expand}
    Z_{T[\mathrm{SU}(2)]}[W_1,W_2;\nu]\xrightarrow{b^2\rightarrow 0}\xi \int \frac{dZ}{\sqrt{2\pi \hbar}}\exp\biggr[\frac{1}{\hbar}\mathcal{W}^{T[\mathrm{SU}(2)]}_0+\mathcal{W}^{T[\mathrm{SU}(2)]}_1+\mathcal{O}(\hbar)\biggr]
\end{align}
where 
\begin{align}
    \begin{split}
        \mathcal{W}_0^{T[\mathrm{SU}(2)]}(Z,W_1,W_2;\nu)&\equiv Z^2+2ZW_2+W_1^2+\tilde{\mathrm{Li}}_2(e^{(\nu-1)\pi i})+\sum_P \tilde{\mathrm{Li}}_2(e^{P-\frac{(\nu+1)\pi i}{2}})\;,\\
    \mathcal{W}^{T[\mathrm{SU}(2)]}_1(Z,W_1;\nu)\phantom{,W_2}&\equiv\frac{\nu}{2}\tilde{\mathrm{Li}}_1(e^{(\nu-1)\pi i})+\frac{1-\nu}{4}\sum_P \tilde{\mathrm{Li}}_1(e^{P-\frac{(\nu+1)\pi i}{2}})
    \end{split}
\end{align}
for \(P=\lbrace \epsilon_1 Z + \epsilon_2 W_1:\epsilon_i=\pm1\rbrace
\). Here, $\tilde{\mathrm{Li}}_{1,2}$ are polylogarithm functions whose branch cuts are modified to ensure a physically correct singularity structure. See appendix~\ref{Ap: QDL} for their definitions. However, for the computations in this paper, we can simply treat them as ordinary $\mathrm{Li}_{1,2}$. We have added comments on the related steps.

\subsection{$D(\vec{k})|_A$ theory}
\subsubsection{Squashed 3-sphere partition function}
From the $T[\rm{SU}(2)]$ partition function \eqref{eq: tsu2 partition function}, the squashed 3-sphere partition function for the $D(\vec{k})$ theory is evaluated as $(\hbar\equiv 2\pi ib^2)$
\begin{align}
    \begin{split}
        Z^{S^3_b}_{D(\vec{k})}[\nu]=\xi\int \biggr[\prod_{i=1}^n\frac{dW_i}{\sqrt{2\pi\hbar}} \Delta_{\mathfrak{su}(2)}^{S^3_b}(W_i)e^{k_i W_i^2/\hbar}\biggr]\biggr[\prod_{i=1}^{n-1}Z_{T[\mathrm{SU}(2)]}[ W_i,W_{i+1};\nu]\biggr]\;.
    \end{split}
\end{align}
Here, $\Delta^{S^3_b}_{\mathfrak{su}(2)}$ is the integration measure for the $\mathfrak{su}(2)$-valued vector multiplet\cite{Hama:2011ea}
\begin{align}
    \Delta^{S^3_b}_{\mathfrak{su}(2)}(W)\equiv2 \sinh W\sinh\frac{2\pi i W}{\hbar}
\end{align}
and $\xi$ is the subtle phase factor. In the $b^2\rightarrow 0$ limit, we can expand $Z^{S^3_b}_{D(\vec{k})}[\nu]$ as\footnote{We assumed that one of the two contributions from the $\sinh\frac{2\pi iW}{\hbar}$ is suppressed in the limit $b^2\rightarrow 0$.}
\label{eq: D(k) expand}
\begin{align}
    \xi\int \frac{1}{(2\pi \hbar)^{n-1/2}} \prod_{i=1}^n dW_i \prod _{i=1}^{n-1}dZ_i \exp\biggr[\frac{1}{\hbar}\mathcal{W}_0^{D(\vec{k})}+\mathcal{W}_1^{D(\vec{k})}+\mathcal{O}(\hbar)\biggr]
\end{align}
where 
\begin{align}
\label{eq: D(k) Ws}
    \begin{split}
        \mathcal{W}_0^{D(\vec{k})}(\vec{Z},\vec{W};\nu)&=\sum_{i=1}^n k_i W_i^2+\sum_{i=1}^{n-1}\mathcal{W}_0^{T[\mathrm{SU}(2)]}(Z_{i},W_i,W_{i+1};\nu)\;,\\
    \mathcal{W}_1^{D(\vec{k})}(\vec{Z},\vec{W};\nu)&=\sum_{i=1}^n\log( \sinh W_i)+\sum_{i=1}^{n-1}\mathcal{W}_1^{T[\mathrm{SU}(2)]}(Z_i,W_i;\nu)\;.
    \end{split}
\end{align}
Note that contributions of the adjoint chirals in the $T[\mathrm{SU}(2)]$ segments to $\CW_1^{D(\vec{k})}$
\begin{align}
    \prod_{k=1}^{n-1}\psi_{\hbar}\left((1-\n)\bigr(\pi i+\frac{\hbar}{2}\bigr)\right)\xrightarrow{b^2\rightarrow 0+}\exp\biggr[\sum_{k=1}^{n-1}\biggr(\frac{\tilde{\mathrm{Li}}_2(e^{(\nu-1)\pi i})}{\hbar}+\frac{\nu}{2}\tilde{\mathrm{Li}}_1(e^{(\nu-1)\pi i})+\mathcal{O}(\hbar)\biggr)\biggr]
\end{align}
becomes singular when $\nu=-1$ due to the presence of $\tilde{\mathrm{Li}}_1(e^{(\nu-1)\pi i})$ term.

\subsubsection{Bethe equation and asymptotic solutions} From the $b^2\rightarrow0$ expansion \eqref{eq: D(k) Ws}, we get the Bethe equation for the $D(\vec{k})$ theory
\begin{align}
    \label{eq: exp BE}
    \lbrace1=e^{\partial \mathcal{W}_0^{D(\vec{k})}}\rbrace\xrightarrow[z_i = e^{Z_i}, \,w_i = e^{W_i}]{\eta=e^{-\frac{1}{2}(\nu+1)\pi i}}\begin{cases}
        1=w_{i+1}^2\frac{(z_i w_i-\eta)(z_i-w_i\eta)}{(w_i-z_i \eta)(1-z_i w_i \eta)}&(i=1,\cdots,n-1)\\
        1=w_{i}^{2k_i} z_{i-1}^2  \frac{(z_i w_i-\eta)(w_i-z_i\eta)}{(z_i-w_i \eta)(1-z_i w_i\eta)}&(i=1,\cdots,n-1)\\
        1=w_n^{2k_n}z_{n-1}^2&
    \end{cases}
\end{align}
where $z_0\equiv 1$. Consider the double-scaling ansatz
\begin{align} \label{eq: Ansatz}
    \n=-1+\e\;,\quad Z_i=(-1)^{G_i}W_i+f_i\e+\CO(\e^2)\;,\quad G_i\in \mathbb{Z}_2\text{ for }i=1,2,\cdots,n-1\;.
\end{align}
Substituting ansatz into un-exponentiated Bethe equation $\pd\CW_0\in2\pi i\mathbb{Z}$, we get
\begin{align}
\label{eq: BE original}
    \lbrace\partial \mathcal{W}_0^{D(\vec
    k)}\in2\pi i \mathbb{Z}\rbrace\to\begin{cases}
        2W_{i+1}-\zeta_i+\mathcal{O}(\epsilon)&\in\pi i (2\mathbb{Z}+1) \\
        2k_i W_i + 2(-1)^{G_{i-1}}W_{i-1} +(-1)^{G_i}\zeta_i+\mathcal{O}(\epsilon)&\in \pi i (2\mathbb{Z}+1)\\
        2k_n W_n + 2(-1)^{G_{n-1}}W_{n-1}&\in2\pi i\mathbb{Z}
        \end{cases}
\end{align}
where \(i=1,2,\cdots,n-1\), $\zeta_i\equiv\log\frac{\pi i-2f_i}{\pi i+2f_i}$ and \(W_0\equiv0\). After adding and subtracting \(\partial_{Z_i}\) and \(\partial_{W_i}\) equations and take \(\epsilon\to 0\) limit, we get
\begin{align}
   \mathrm{K}\vec{W}^{\vec{G}} \in \pi i \mathbb{Z}^n, \quad \tilde{\mathrm{K}}\vec{W}^{\vec{G}} + \vec{\zeta}^{\vec{G}}\in \pi i \mathbb{Z}^n
\end{align}
where
\begin{align}
    \mathrm{K}[\vec{k}]\equiv \begin{pmatrix}
        k_1 & 1 & 0 & 0&\cdots &0&0&0 \\
        1 & k_2 & 1 & 0 & \cdots&0&0&0\\
        0&1&k_3  & 1 & \cdots &0&0&0\\
        0 &0 & 1 & k_4  & \cdots&0&0&0\\
        \vdots&\vdots&\vdots&\vdots&\ddots&\vdots&\vdots&\vdots\\
        0&0&0&0&\cdots&k_{n-2}&1&0\\
        0&0&0&0&\cdots&1&k_{n-1}&1\\
        0&0&0&0&\cdots&0&1&k_n\\    
    \end{pmatrix}\;,\quad \tilde{\mathrm{K}}[\vec{k}]\equiv \begin{pmatrix} 
        k_1 & -1 & 0 & 0&\cdots &0&0&0 \\
        1 & k_2  & -1 & 0 & \cdots&0&0&0\\
        0&1&k_3 & -1 & \cdots &0&0&0\\
        0 &0 & 1 & k_4 & \cdots&0&0&0\\
        \vdots&\vdots&\vdots&\vdots&\ddots&\vdots&\vdots&\vdots\\
        0&0&0&0&\cdots&k_{n-2}&-1&0\\
        0&0&0&0&\cdots&1&k_{n-1}&-1\\
        0&0&0&0&\cdots&0&1&k_n\\    
    \end{pmatrix}
\end{align}
and \(W_i^{\vec{G}}\equiv W_i\prod_{j=1}^{i-1}(-1)^{G_j}\), \(\zeta_i^{\vec{G}} \equiv \zeta_i\prod_{j=1}^{i}(-1)^{G_j}\). By accounting all possible choices of $\mathbb{Z}^n$ in the RHS of the first equation, we find ${\rm BV}^{D(\vec{k})}$ shown in \eqref{eq: D(k) Before WQ BV}. Note that from \eqref{eq: BE original}, we can determine $f_i$ by taking exponential on both sides of the first equation
\begin{align}
        e^{2W_{i+1}}\frac{\pi i+2f_i}{\pi i-2f_i}=-1\longrightarrow f_i=-\frac{\pi i}{2}\coth W_{i+1}\;.
\end{align}
It serves as a \textit{consistency equation} for the ansatz \eqref{eq: Ansatz}. $W_{i+1}\in\pi i\mathbb{Z}$ is \textit{forbidden} since $f_{i}$ is not determined.

\subsubsection{Handle-gluing from asymptotic solutions}
For the solutions of Bethe equation before taking the Weyl quotient \eqref{eq: D(k) Before WQ BV}, let's calculate associated Handle-gluings.
\paragraph{Determinant factor} First, consider the determinant factor. In \(\epsilon\to 0\) limit, we have
\begin{align}
    \begin{split}
        \pd_{Z_i}^2\CW^{D(\vec{k})}_0&=\frac{1}{\e}\frac{-4\pi i}{\pi^2+4f_i^2}+\tau_i+\CO(\e)\;,\quad \pd_{Z_{i}}\pd_{W_{i+1}}\CW_0^{D(\vec{k})}=2\;,
        \\\pd_{W_i}^2\CW_0^{D(\vec{k})}&=\frac{1}{\e}\frac{-4\pi i}{\pi^2+4f_i^2}+2k_i+\tau_i+\CO(\e)\;,\quad \pd^2_{W_n}\CW_0^{D(\vec{k})}=2k_n\;,
        \\\pd_{Z_i}\pd_{W_i}\CW_0^{D(\vec{k})}&=(-1)^{G_{i}+1}\biggr[\frac{1}{\e}\frac{-4\pi i}{\pi^2+4f_i^2}+\tau_i\biggr]+\CO(\e)\;,\quad(\text{Otherwise})=0\;.
    \end{split}
\end{align}
Here, $\tau_i$ is a term depends on the second order expansion coefficient but it doesn't contribute to the result. The determinant of hessian is evaluated as
\begin{align}
        \det\left[-\partial^2\mathcal{W}_0^{D(\vec{k})|_A}\right]=\lim_{\epsilon\to 0} \det\left[-\partial^2 \mathcal{W}_0^{D(\vec{k})}\right] &= \lim_{\epsilon\to 0} \left[-\frac{2^n p}{\epsilon^{n-1}}\prod_{i=1}^{n-1}\frac{-4\pi i }{\pi^2 + 4f_i^2}\right]\;. 
\end{align}
\paragraph{Exponential factor} Independent of the choice of $\vec{G}$, $\CW_1^{T[{\rm SU}(2)]}$ becomes
\begin{align}
        e^{\mathcal{W}_1^{T[\textrm{SU(2)}]|_A}}&=\pm\lim_{\epsilon\to0}\sqrt{\frac{\pi i }{-\epsilon(4f_i^2 + \pi^2) \sinh^2 W_i}}
\end{align}
in the $\epsilon\rightarrow 0$ limit and the exponential factor is evaluated as
\begin{align}
        e^{\mathcal{W}_1^{D(\vec{k})|_A}}&=\pm \lim_{\epsilon\to0}\frac{\sinh W_n}{2^{n-1}\epsilon^{(n-1)/2}}\prod_{i=1}^{n-1}\sqrt{\frac{-4\pi i}{\pi^2 + 4f_i^2}}\;.
\end{align}

\paragraph{Handle-gluing} Combining two results, we get
\begin{align}
    \mathcal{H}= \frac{e^{i\delta}}{|\mathbf{Weyl}|^2 }\det\left[-\partial^2 \mathcal{W}_0^{D(\vec{k})|_A}\right]e^{-2\mathcal{W}_1^{D(\vec{k})|_A}} = -\frac{e^{i\delta}2^{n-2}p}{\sinh^2 W_n}\;.
\end{align}
Finally, we can determine $e^{i\delta}$ by requiring
\begin{align}
    \begin{split}
        \sum_{\frac{\mathrm{BV}^{D(\vec{k})}}{(\mathbf{Weyl})}}\mathcal{H}^{-1}&=\frac{e^{-i\delta }}{2^{n-2}p}\times\sum_{\vec{n}}\sum_{\alpha=1}^{|p|-1}\sin^{2}\biggr(\pi\biggr(n_n+\frac{\pi }{p}\alpha(-1)^{n-1}\biggr)(-1)^{G_n}\prod_{j=1}^{n-1}(-1)^{G_j}\biggr)
    \\&=\frac{e^{-i\delta}}{2^{n-2}p}\times \sum_{\vec{n}}\sum_{\alpha=1}^{|p|-1}\sin^2\biggr(\frac{\pi }{p}\alpha\biggr)=\frac{e^{-i\delta}}{2^{n-2}p}\times 2^{n-1}\times \frac{|p|}{2}
    \\&={\rm sgn}(p)e^{-i\delta}\longrightarrow{\rm sgn}(p)e^{-i\delta}=1\;.
    \end{split}
\end{align}
So the Handle-gluing is independent of $\vec{n}$ and $\vec{G}$:
\begin{align}
    \CH=-\frac{2^{n-2}|p|}{\sinh^2W_n}=2^{n-2}|p|\sin^{-2}\biggr(\frac{\pi}{2}\a\biggr)\;.
\end{align}

\subsubsection{Fibering from asymptotic solutions}
For the solutions of Bethe equation before taking the Weyl quotient \eqref{eq: D(k) Before WQ BV}, let's calculate modified $\mathcal{W}_0^{D(\vec{k})}$
\begin{align}
    \mod \mathcal{W}_0^{D(\vec{k})} \equiv \mathcal{W}^{D(\vec{k})}_0-\vec{Z}\cdot \partial_{\vec{Z}}\mathcal{W}^{D(\vec{k})}_0-\vec{W}\cdot \partial_{\vec{W}}\mathcal{W}^{D(\vec{k})}_0\;.
\end{align}
Note that the linear modification terms in ${\tilde{\rm Li}}_2$ and the modification term in ${\tilde{\rm Li}}_1$ are mutually canceled. Also note that it is enough to do calculations modulo $4\pi^2$ since the fibering is defined as $e^{-\frac{{\rm mod}\CW_0}{2\pi i}}$. So we can use ordinary ${\rm Li}_{1,2}$. Using the inversion formula \(\textrm{Li}_2(z)+\textrm{Li}_2(z^{-1}) = -\frac{\pi^2}{6}- \frac{1}{2}(\log (-z))^2\), we get
\begin{align}
    \begin{split}
        \mathcal{W}_0^{D(\vec{k})|_A} = \sum_{i=1}^{n}k_i W_i^2 + \sum_{i=1}^{n-1}\biggr[ 2Z_i W_{i+1}+\frac{5\pi^2}{6}+2\pi i W_i + \sum_{\pm}2\pi i \biggr\lfloor \frac{W_i \pm Z_i}{2\pi i }\biggr\rfloor(W_i \pm Z_i) \biggr]
    \end{split}
\end{align}
and the modified \(\mathcal{W}_0^{D(\vec{k})|_A}\) is evaluated as
\begin{align}
        \mod \mathcal{W}_0^{D(\vec{k})|_A} =\frac{5(n-1)\pi^2}{6}-\sum_{i=1}^{n}k_i W_i^2-2\sum_{i=1}^{n-1}(-1)^{G_i}W_i W_{i+1}\;.
\end{align}
This result is independent of choice of $\vec{G}$ since $(W_i^{\vec{n},\vec{G}})^2=(W_i^{\vec{n},\vec{G}=\vec{0}})^2$ and
\begin{align}\begin{split}
(-1)^{G_i}W_i^{\vec{n},\vec{G}} W_{i+1}^{\vec{n},\vec{G}}&=(-1)^{G_i}\biggr[(-1)^{2G_n}\prod_{j=1}^{i-1}(-1)^{G_j}\prod_{k=1}^{i}(-1)^{G_k}\biggr]W_i^{\vec{n},\vec{G}=\vec{0}}W_{i+1}^{\vec{n},\vec{G}=\vec{0}}
\\&=\biggr[\prod_{k=1}^{i}(-1)^{k}\biggr]^2W_i^{\vec{n},\vec{G}=\vec{0}}W_{i+1}^{\vec{n},\vec{G}=\vec{0}}
\\&=W_i^{\vec{n},\vec{G}=\vec{0}}W_{i+1}^{\vec{n},\vec{G}=\vec{0}}\;.
\end{split}\end{align}
So it is enough to evaluate when $\vec{G}=\vec{0}$. Using the following properties
\begin{align}
    k_i q_i^2 =\begin{cases}
        q_1(p-q_2) &(i=1)
        \\-q_i(q_{i-1}+q_{i+1}) &(i=2,\cdots,n-1)
        \\ -q_{n-1}q_n &(i=n)
    \end{cases}
\end{align}
we get
\begin{align}
\mod \mathcal{W}_0^{D(\vec{k})|_A}=\frac{5}{6}(n-1)\pi^2+\frac{\pi^2}{p}q\alpha^2+\pi^2\biggr[\sum_{i=1}^n k_i n_i^2+2\sum_{i=1}^{n-1}n_i n_{i+1}\biggr]+2 \pi^2 n_1\a
\end{align}
and the Fibering is evaluated as
\begin{align}
    \CF=e^{-\frac{{\rm mod}\CW^{D(\vec{k})|_A}_0}{2\pi i}}=\CF_0\exp\biggr(\frac{\pi i}{2p}q\a^2+\pi in_1\a+\frac{\pi i}{2}\vec{n}^T{\rm K}[\vec{k}]\vec{n}\biggr)\quad(\CF_0\equiv e^{\frac{5}{12}(n-1)\pi i})\;,
\end{align}
independent of $\vec{G}$(\textbf{Weyl}).

\subsubsection{Factorization of $\HF$ data}
\paragraph{When $p$ is odd}
Let's split $\{\a\}=\{1,2,\cdots,|p|-1\}\equiv \{\a_0\}\cup\{\a_1\}$ where
\begin{align}
    \{\a_0\}\equiv\{2,4,6,\cdots,|p|-1\}\text{ and }\{\a_1\}\equiv\{1,3,5,\cdots,|p|-2\}\;.
\end{align}
Then $\{\a_1\}=\bigr\{|p|-\a_0\;\bigr|\;\a_0\in\{\a_0\}\bigr\}$ and for every $\a_0$,
\begin{align}
    \CF(|p|-\a_0,\vec{n})=e^{\pi i(\frac{pq}{2}+n_1)}\CF(\a_0,\vec{n})\;.
\end{align}
So we can find $2^n$ sets of modular data which have identical Handle-gluings and fiberings are differ only in constant phase shift, labeled by $t\in\mathbb{Z}_2$ and $\vec{n}\in(\mathbb{Z}^{n-1}_2,0)$:
\begin{align}
    \biggr\{\biggr(2^{n-2}|p|\sin^{-2}\bigr(\frac{\pi \a_0}{p}\bigr),\CF_0e^{\pi i(\frac{pq}{2}+n_1)t}\exp\bigr(\frac{\pi i}{2p}q\a_0^2+\frac{\pi i}{2}\vec{n}^T{\rm K}[\vec{k}]\vec{n}\bigr)\biggr)\;\biggr|\;\a_0\in\{\a_0\}\biggr\}\;.
\end{align}
So $\HF$ data of the $D(\vec{k})|_A$ theory is factorized as
\begin{align}
    \HF^{D(\vec{k})}=\HF^{\CD(p,q)}\otimes \HF^{\TFTk}
\end{align}
where
\begin{align}
\begin{split}
        \mathbf{HF}^{\mathcal{D}(p,q)}&\equiv\biggr\{\biggr(\frac{|p|}{4}\sin^{-2}\bigr(\frac{2\pi a}{p}\bigr),\mathcal{F}_0\exp\bigr(\frac{2\pi i}{p}q a^2\bigr)\biggr)\;\biggr|\;a\in\bigr\{1,2,\cdots,\frac{|p|-1}{2}\bigr\}\biggr\}\;,
        \\\mathbf{HF}^{\TFTk}&\equiv \biggr\{\biggr(2^{n},e^{\pi i(\frac{pq}{2}+n_1)t}\exp\bigr(\frac{\pi i}{2}\vec{n}^T \mathrm{K}[\vec{k}]\vec{n}\bigr)\biggr)\;\biggr|\;t\in\mathbb{Z}_2,\;\vec{n}\in(\mathbb{Z}^{n-1}_2,0)\biggr\}\;.
\end{split}
\end{align}
Note that we introduce the new label $2a\equiv \a_0$.

\begin{figure}[h]
\centering
\includegraphics[width=.80\textwidth]{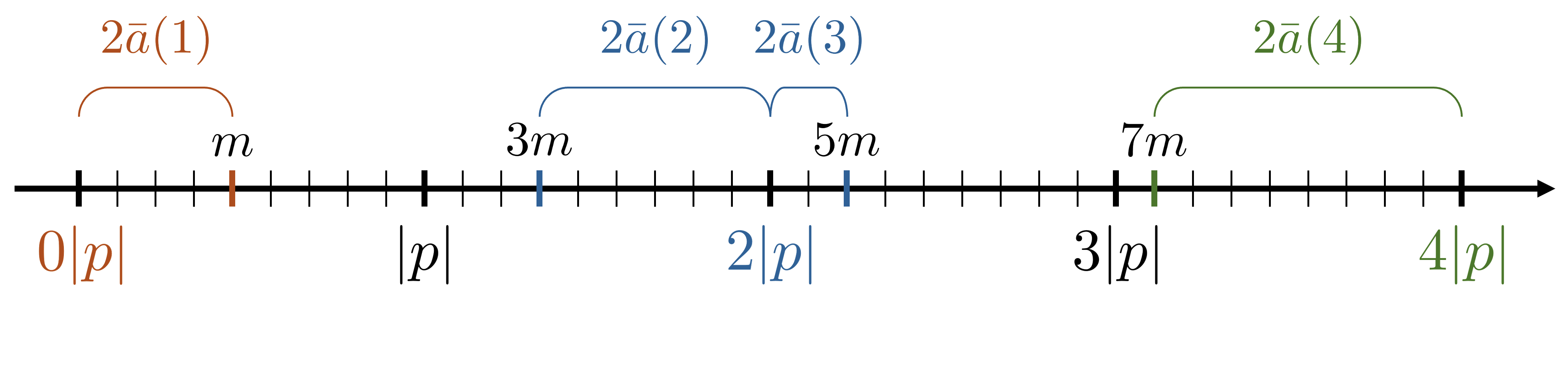}
\captionof{figure}{Example of $\bar{a}(a)$ when $|p|=9$, $m=4$.}
\label{fig: baranumbering}
\end{figure}

Both $\HF$ data satisfy the condition \eqref{eq: HF data relation}. There is a unique even $m\in\{2,4,6,\cdots,|p|-1\}$ and $s\in\{\pm 1\}$ satisfies $|q|m\in|p|\mathbb{N}_0+s$ and we can check
\begin{align}
    \biggr|\sum_{\HF^{\CD(p,q)}}\CH^{-1}\CF\biggr|=\bigr|\CH(a=\frac{m}{2})\bigr|^{-\frac{1}{2}}\;,\quad \biggr|\sum_{\HF^{\TFTk}}\CH^{-1}\CF\biggr|=\bigr|\CH_{\TFTk}=2^{n}\bigr|^{-\frac{1}{2}}\;.
\end{align}
Furthermore, we can identify $\HF^{\CD(p,q)}$ with $\HF$ data of the WRT TQFT. Let's consider relabeling of $\HF^{\CD(p,q)}$
\begin{align} \label{eq: bar a}
    \bar{a}(a)\equiv\biggr|\frac{p}{2\pi}{\rm Log}\;e^{\frac{\pi i}{p}m(2a-1)}\biggr|\text{ for }a\in\bigr\{1,2,\cdots,\frac{|p|-1}{2}\bigr\}\;.
\end{align}
Note that $2\bar{a}(a)$ is same with the distance between $m(2a-1)$ and the nearest point in $2|p|\mathbb{N}_0$. See figure~\ref{fig: baranumbering} below to get the picture.

\noindent Then we can check an identification up to the phase factor $\xi$
\begin{align}
    \biggr(\frac{|p|}{4}\sin^{-2}\bigr(\frac{2\pi}{p}\bar{a}(a)\bigr),\CF_0\exp\bigr(\frac{2\pi i}{p}q(\bar{a}(a))^2\bigr)\biggr)=\bigr((\bar{S}_{1a})^{-2},\xi\bar{T}_{aa}\bigr)
\end{align}
where $\bar{S}$ and $\bar{T}$ are modular matrices of
\begin{align} \label{eq: odd p WRT identification from HF data}
    \begin{cases}
        {\rm WRT}^m_{k=|p|-2}&(s\times{\rm sgn}(pq)>0)\\
        {\rm WRT}^{2|p|-m}_{k=|p|-2}&(s\times {\rm sgn}(pq)<0)
    \end{cases}\;.
\end{align}

\paragraph{When $p$ is even} Since $q$ is odd, we find
\begin{align}
\begin{split}
    &\CF(|p|-\a,\vec{n})\bigr|_{n_1=0}=e^{\pi i\frac{pq}{2}}\CF(\a,\vec{n})\bigr|_{n_1=1}
    \\&\text{where }\CF(\a,\vec{n})\bigr|_{n_1=1}=\CF_0\exp\biggr(\frac{\pi i}{2p}q\a^2+\pi i\a\biggr)\exp\biggr(\frac{\pi i}{2}\vec{n}^T{\rm K}[\vec{k}]\vec{n}\biggr)\;.
\end{split}
\end{align}
From this, we can find $2^{n-1}$ set of modular data which have identical Handle-gluings and fiberings are differ only in constant phase shift labeled by $\vec{n}\in(\mathbb{Z}_2^{n-1},0)$:
\begin{align}
    \biggr\{\biggr(2^{n-2}|p|\sin^{-2}\bigr(\frac{\pi \a}{p}\bigr),\CF_0e^{-\pi i\frac{pq}{2}n_1}\exp\bigr(\frac{\pi i}{2p}q\a^2+\frac{\pi i}{2}\vec{n}^T{\rm K}[\vec{k}]\vec{n}\bigr)\biggr)\;\biggr|\;\a\in\{1,2,\cdots,|p|-1\}\biggr\}\;.
\end{align}
So $\HF$ data of the $D(\vec{k})|_A$ theory is factorized as
\begin{align}
    \HF^{D(\vec{k})}=\HF^{\CD(p,q)}\otimes \HF^{\TFTk}
\end{align}
where
\begin{align}
\begin{split}
        \mathbf{HF}^{\mathcal{D}(p,q)}&=\biggr\{\biggr(\frac{|p|}{2}\sin^{-2}\bigr(\frac{\pi \a}{p}\bigr),\mathcal{F}_0\exp\bigr(\frac{\pi i}{2p}q \a^2\bigr)\biggr)\;\biggr|\;\a\in\{1,2,\cdots,|p|-1\}\biggr\}\;,
        \\\mathbf{HF}^{\TFTk}&= \biggr\{\biggr(2^{n-1},e^{-\pi i\frac{pq}{2}n_1}\exp\bigr(\frac{\pi i}{2}\vec{n}^T \mathrm{K}[\vec{k}]\vec{n}\bigr)\biggr)\;\biggr|\;\vec{n}\in(\mathbb{Z}^{n-1}_2,0)\biggr\}\;.
\end{split}
\end{align}

\begin{figure}[h]
\centering
\includegraphics[width=.80\textwidth]{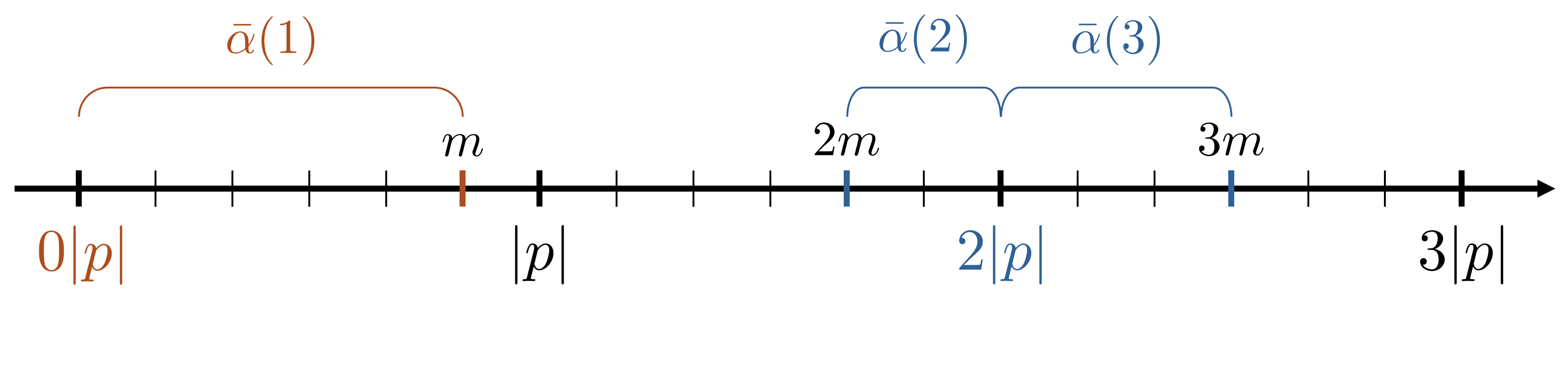}
\captionof{figure}{Example of $\bar{\alpha}(\alpha)$ when $|p|=6$, $m=5$.}
\label{fig: baralphanumbering}
\end{figure}

Both $\HF$ data satisfy the condition \eqref{eq: HF data relation}. There is a unique odd $m\in\{1,3,5,\cdots,|p|-1\}$, $N\in\mathbb{N}_0$ and $s\in\{\pm 1\}$ satisfy $|q|m=2|p|N+s$ and we can check
\begin{align}
    \biggr|\sum_{\HF^{\CD(p,q)}}\CH^{-1}\CF\biggr|=\bigr|\CH(\a=m\text{ or }|p|-m)\bigr|^{-\frac{1}{2}}\;,\quad \biggr|\sum_{\HF^{\TFTk}}\CH^{-1}\CF\biggr|=\bigr|\CH_{\TFTk}=2^{n-1}\bigr|^{-\frac{1}{2}}\;.
\end{align}
Furthermore, we can identify $\HF^{\CD(p,q)}$ with $\HF$ data of the WRT TQFT. Let's consider relabeling of $\HF^{\CD(p,q)}$
\begin{align} \label{eq: bar alpha}
    \bar{\a}(\a)\equiv\biggr|\frac{p}{\pi}{\rm Log}\;e^{\frac{\pi i}{p}m\a}\biggr|\text{ for }\a\in\{1,2,\cdots,|p|-1\}\;.
\end{align}
Note that $\bar{\a}(\a)$ is same with the distance between $m\a$ and the nearest point in $2|p|\mathbb{N}_0$. See figure~\ref{fig: baralphanumbering} below to get the picture.

\noindent Then we can check an identification up to the phase factor $\xi$
\begin{align}
    \biggr(\frac{|p|}{2}\sin^{-2}\bigr(\frac{\pi}{2}\bar{\a}(\a)\bigr),\CF_0\exp\bigr(\frac{\pi i}{2p}q(\bar{\a}(\a))^2\bigr)\biggr)=\bigr((S_{1\a})^{-2},\xi T_{\a\a}\bigr)
\end{align}
where $S$ and $T$ are modular matrices of
\begin{align} \label{eq: even p WRT identification from HF data ver 1}
    \begin{cases}
        {\rm WRT}^{m}_{k=|p|-2}&(s\times{\rm sgn}(pq)>0,(-1)^N>0)\\
        {\rm WRT}^{4|p|-m}_{k=|p|-2}&(s\times{\rm sgn}(pq)<0,(-1)^N>0)\\
        {\rm WRT}^{2|p|+m}_{k=|p|-2}&(s\times{\rm sgn}(pq)>0,(-1)^N<0)\\
        {\rm WRT}^{2|p|-m}_{k=|p|-2}&(s\times{\rm sgn}(pq)<0,(-1)^N<0)\\
    \end{cases}\;.
\end{align}
However, \eqref{eq: even p WRT identification from HF data ver 1} is NOT a unique identification. We can choose
\begin{align}
    \bar{\bar{\a}}(\a)\equiv|p|-\bar{\a}(\a)\text{ for }\a\in\{1,2,\cdots,|p|-1\}
\end{align}
alternatively and check the other identifications
\begin{align} \label{eq: even p WRT identification from HF data ver 2}
    \begin{cases}
        {\rm WRT}^{2|p|+m}_{k=|p|-2}&(s\times{\rm sgn}(pq)>0,(-1)^N>0)\\
        {\rm WRT}^{2|p|-m}_{k=|p|-2}&(s\times{\rm sgn}(pq)<0,(-1)^N>0)\\
        {\rm WRT}^{m}_{k=|p|-2}&(s\times{\rm sgn}(pq)>0,(-1)^N<0)\\
        {\rm WRT}^{4|p|-m}_{k=|p|-2}&(s\times{\rm sgn}(pq)<0,(-1)^N<0)\\
    \end{cases}\;.
\end{align}

\section{Candidates of simple objects}
\label{Ap: Simple obejcts}
Let's denote the (Supersymmetric) Wilson line in the $s$-th symmetric representation of the ${\rm SU}(2)^{i=1,2,\cdots,n}$ following the curve $\CC$ as $W_{\CC}^{(i,s)}$. We can also consider the Wilson line charged under the ${\rm U}(1)^{i}$ with the charge $Q_i\in\mathbb{Z}$, following the curve $\mathcal{C}$. We denote it as $\tilde{W}^{(i,Q_i)}_{\mathcal{C}}$. On the squashed 3-sphere background, there are the two possible curves $\CC_\pm$ allowed in the localization as shown in figure~\ref{fig: SUSY lines in S3b}. For each curve, Wilson lines are evaluated as\cite{Kapustin:2009kz,Closset:2018ghr}
\begin{align}
\label{eq: Wilson lines}
    W^{(i,s)}_{\mathcal{C}_\pm}=\sum_{j=0}^{s}e^{(s-2j)\frac{W_i}{b}\times b^{\pm1}}=\frac{\sinh\bigr((s+1)\frac{W_i}{b}\times b^{\pm1}\bigr)}{\sinh (\frac{W_i}{b}\times b^{\pm1})}\;,\quad \tilde{W}^{(i,Q_i)}_{\mathcal{C}_\pm}=e^{Q_i\frac{Z_i}{b}\times b^{\pm 1}}\;.
\end{align}
Using the curve $\mathcal{C}_+$, we can test the IR fate of the Wilson line through the criteria \eqref{eq: Simple crit TTI}.
\subsection{When $p$ is odd}
Motivated by the fact that the decoupled TQFT of the $D(\vec{k})|_A$ theory is expected to originate from the $\mathbb{Z}_2$ 1-form symmetry associated with each ${\rm SU}(2)^{i=1,2,\cdots,n}$\cite{Hsin:2018vcg,Gang:2024wxz} and it seems to be not altering the value of Handle-gluings(We can infer this from the fact that all Handle-gluings of $\HF^{\TFTk}$ have the same value), it is natural to consider the following $(|p|-1)\times 2^{n-1}$ objects
\begin{align} \label{eq: odd p Simples}
    \CO[a,t,\vec{n}]\equiv\biggr[\bigotimes_{i=1}^{n-1}[W^{(i,|p|-2)}]^{n_i}\biggr]\otimes[W^{(n,|p|-2)}]^{t}\otimes W^{(n,2(a-1))}
\end{align}
labeled by $a\in\bigr\{1,2,\cdots,\frac{|p|-1}{2}\bigr\}$, $t\in\mathbb{Z}_2$ and $\vec{n}\in(\mathbb{Z}^{n-1}_2,0)$. \eqref{eq: odd p Simples} is evaluated as
\begin{align} \label{eq: odd p SO values}
    \CO_{\CC_+}[a,t,\vec{n}]\bigr|_{V(\a',\vec{n}')}=(-1)^{\vec{n}\cdot\vec{n}'+\a'\vec{q}\cdot\vec{n}+\sum_{i=1}^{n-1}n_i+(\a'+1)t}\times\frac{\sin\bigr(\frac{\pi}{p}(2a-1)\a'\bigr)}{\sin\bigr(\frac{\pi}{p}\a'\bigr)}
\end{align}
for the Bethe vacuum $V(\a',\vec{n}')$ and form the maximally independent set
\begin{align}
    \sum_{V\in\frac{{\rm BV}^{D(\vec{k})}}{(\textbf{Weyl})}}\CH^{-1}\CO_{\CC_+}[a_1,t_1,\vec{n}_1]\bigr(\CO_{\CC_+}[a_2,t_2,\vec{n}_2]\bigr)^*\biggr|_V=\d_{[a_1,t_1,\vec{n}_1],[a_2,t_2,\vec{n}_2]}\;.
\end{align}

Let us consider an alternative set constructed by replacing some of ${\rm SU}(2)^{i=1,2,\cdots,n-1}$ Wilson lines to ${\rm U}(1)^{i=1,2,\cdots,n-1}$ Wilson lines
\begin{align} \label{eq: odd p alt Simples}
    \CO^{(\vec{L},\vec{Q})}[a,t,\vec{n}]\equiv \biggr[\bigotimes_{i=1}^{n -1} [W^{(i,|p|-2)}]^{(1-L_i)n_i}\biggr]\otimes\biggr[\bigotimes_{i=1}^{n-1}[\tilde{W}^{(i,Q_i)}]^{L_in_i}\biggr]\otimes [W^{(n,|p|-2)}]^t\otimes W^{(n,2(a-1))}
\end{align}
labeled by $a\in\bigr\{1,2,\cdots,\frac{|p|-1}{2}\bigr\}$, $t\in\mathbb{Z}_2$, $\vec{n}\in(\mathbb{Z}^{n-1}_2,0)$, $\vec{L}\in\mathbb{Z}_2^{n-1}$ and $Q_{i=,1,2,\cdots,n-1}\in2\mathbb{Z}-1$. \eqref{eq: odd p alt Simples} is evaluated as
\begin{align}
    \begin{split}
        \CO_{\CC_+}^{(\vec{L},\vec{Q})}[a,t,\vec{n}]\bigr|_{V(\a',\vec{n}')}=&e^{\frac{\pi i}{p}\a'\sum_{i=1}^{n-1}[Q_iL_iq_in_i]}
        \\&\times(-1)^{\vec{n}\cdot\vec{n}'+\sum_{i=1}^{n-1}[(1-L_i)(n_i+\a'q_in_i)]+(\a'+1)t}\times\frac{\sin\bigr(\frac{\pi}{p}(2a-1)\a'\bigr)}{\sin\bigr(\frac{\pi}{p}\a'\bigr)}
    \end{split}
\end{align}
for the Bethe vacuum $V(\a',\vec{n}')$ and we can check it also form the maximally independent set. Note that $\CO^{(\vec{L}=\vec{0},\vec{Q})}=\CO$.

\subsection{When $p$ is even}
For this case, ${\rm SU}(2)^{i=1,2,\cdots,n-1}$ Wilson lines $W^{(i,|p|-2)}$ are not possible candidates for simple objects of the decoupled TQFT of the $D(\vec{k})|_A$ theory. However, ${\rm U}(1)^{i=1,2,\cdots,n-1}$ Wilson lines of odd charges still pass the criteria \eqref{eq: Simple crit TTI}. We can check $(|p|-1)\times 2^{n-1}$ objects
\begin{align} \label{eq: even p Simples}
    \bar{\CO}^{\vec{Q}}[\a,\vec{n}]\equiv \biggr[\bigotimes_{i=1}^{n-1}[\tilde{W}^{(i,Q_i)}]^{n_i}\biggr]\otimes W^{(n,\alpha-1)}
\end{align}
labeled by $\a\in\{1,2,\cdots,|p|-1\}$, $\vec{n}\in(\mathbb{Z}^{n-1}_2,0)$ and $Q_{i=1,2,\cdots,n-1}\in2\mathbb{Z}-1$ are evaluated as
\begin{align} \label{eq: even p SO values}
    \bar{\CO}^{\vec{Q}}_{\CC_+}[\a,\vec{n}]\bigr|_{V(\a',\vec{n}')}=e^{\frac{\pi i}{p}\alpha'\sum_{i=1}^{n-1}[Q_i q_in_i]}(-1)^{\vec{n}\cdot\vec{n}'}\times\frac{\sin\bigr(\frac{\pi}{p}\a\a'\bigr)}{\sin\bigr(\frac{\pi}{p}\a'\bigr)}
\end{align}
and form the maximally independent set
\begin{align}
    \sum_{V\in\frac{{\rm BV}^{D(\vec{k})}}{(\textbf{Weyl})}}\CH^{-1}\bar{\CO}^{\vec{Q}}_{\CC_+}[\a_1,\vec{n}_1]\bigr(\bar{\CO}^{\vec{Q}}_{\CC_+}[\a_2,\vec{n}_2]\bigr)^*\biggr|_{V}=\d_{[\a_1,\vec{n}_1],[\a_2,\vec{n}_2]}\;.
\end{align}

\section{Full modular matrices}
\label{Ap: Full modular matrices}
\subsection{Details of the modular $S$ matrix computation}
\label{Ap: Detail of S matrix computation}
\subsubsection{When $p$ is odd}
Through the 3-sphere partition function analysis, we find that the true Bethe vacuum of the $D(\vec{k})|_A$ theory $V_1$ must be one of the $\a=m$ or $|p|-m$ elements (We still have $\vec{n}$ ambiguity) in a set of Bethe vacua $\frac{{\rm BV}^{D(\vec{k})}}{(\textbf{Weyl})}$ where $m\in\{1,2,\cdots,|p|-1\}$ can be chosen as a unique even integer that satisfies $|q|m\in |p|\mathbb{N}_0\pm 1$. Let's denote
\begin{align} \label{eq: odd p possible V1}
    V_1=V\bigr(|p|t_0+(-1)^{t_0}m,\vec{n}_0\bigr)\in\frac{{\rm BV}^{D(\vec{k})}}{(\textbf{Weyl})}\text{ where }t_0\in\mathbb{Z}_2,\;\vec{n}_0\in(\mathbb{Z}_2^{n-1},0)\;.
\end{align}
Then we can determine the first row of the modular $S$ matrix up to the overall phase factor $\g$ using simple objects \eqref{eq: odd p Simples} as
\begin{align}
    S^{D(\vec{k})}_{1A}=\CO_{\CC_+}[A]\bigr|_{V_1}S^{D(\vec{k})}_{11}\;,\quad S^{D(\vec{k})}_{11}=\g\times 2^{-\frac{n}{2}}\sqrt{\frac{4}{|p|}}\sin\bigr(\frac{\pi }{|p|}m\bigr)\;.
\end{align}
Here, we introduce an index $A\in\{1,2,\cdots,(|p|-1)\times 2^{n-1}\}$ that labels simple objects $\CO[a,t,\vec{n}_i]$ as
\begin{align}
    A=A(a,t,i)\equiv |p|t+(-1)^{t}(2a-1)+|p|(i-1)
\end{align}
where $a\in\bigr\{1,2,\cdots,\frac{|p|-1}{2}\bigr\}$, $t\in\mathbb{Z}_2$ and $\{\vec{n}_{i=1}\equiv\vec{0},\vec{n}_{i=2,\cdots,2^{n-1}}\}=(\mathbb{Z}^{n-1}_2,0)$. In this stage, we can check
\begin{align}
    2^{\frac{n}{2}}\times|S^{D(\vec{k})}_{1A(a=1)}|=|\bar{S}_{11}|=\sqrt{\frac{4}{|p|}}\sin\bigr(\frac{m\pi}{|p|}\bigr)\;,\quad \frac{S^{D(\vec{k})}_{1A}}{S^{D(\vec{k})}_{1A(a=1)}}=\frac{\bar{S}_{1a}}{\bar{S}_{11}}
\end{align}
after some computations using \eqref{eq: odd p SO values}. $\bar{S}$ is a modular $S$ matrix of the WRT TQFT determined from the $\HF$ data analysis. So it is natural to identify
\begin{align}
    a\longrightarrow \text{WRT index}\;,\quad (t,i)\longrightarrow\TFTk\text{ index}\;.
\end{align}
Let's try a Bethe vacuum-Simple object map (See around \eqref{eq: BVSO map}) of form
\begin{align} \label{eq: odd p BVSO map}
    V_{A(a',t',i')}\equiv V\bigr(|p|T'+2(-1)^{T'}\bar{a}(a'),\vec{n}'\bigr)\in\frac{{\rm BV}^{D(\vec{k})}}{(\textbf{Weyl})}\longleftrightarrow\CO[a',t',\vec{n}_{i'}]\;.
\end{align}
Here, $T'\in\mathbb{Z}_2$ and $\vec{n}'\in(\mathbb{Z}_2^{n-1},0)$ are functions of $(t',i')$ that makes $\{(T',\vec{n}')\}$ one-to-one with $\bigr(\mathbb{Z}_2,(\mathbb{Z}^{n-1}_2,0)\bigr)$ under constraints $T'(0,1)=t_0$ and $\vec{n}'(0,1)=\vec{n}_0$. $\bar{a}(a')$ is a rearrangement of $a'$ defined as \eqref{eq: bar a}. Note that $2\bar{a}(1)=m$. So
\begin{align} \label{eq: odd p V1 consistency}
    V_1=V(2\bar{a}(1),\vec{n}_0)=V(m,\vec{n}_0)
\end{align}
and \eqref{eq: odd p V1 consistency} is in agreement with \eqref{eq: odd p possible V1}. Then the full modular matrices of the $D(\vec{k})|_A$ are determined by relations \eqref{eq: S matrix dict} and \eqref{eq: T matrix dict} as
\begin{align}
    S^{D(\vec{k})}_{A(a',t',i')A(a,t,i)}=\bar{S}_{a'a}\times S^{\TFTk}_{(t',i')(t,i)}\;,\quad T^{D(\vec{k})}_{A(a',t',i')A(a,t,i)}=\bar{T}_{a'a}\times T^{\TFTk}_{(t',i')(t,i)}
\end{align}
where
\begin{align}
    \begin{split}
        S^{\TFTk}_{(t',i')(t,i)}&\equiv\gamma'\times 2^{-\frac{n}{2}}(-1)^{\vec{n}_i\cdot\vec{n}'+t_0 t'+tT'+\sum_{k=1}^{n-1}(\vec{n}_i+\vec{n}_{i'})_k+\vec{n}_{i'}\cdot\vec{n}_0+\vec{q}\cdot(t_0\vec{n}_{i'}+T'\vec{n}_i)+t+t'}\;,
        \\T^{\TFTk}_{(t',i')(t,i)}&\equiv\xi'\times\delta_{t't}\delta_{i'i}e^{\pi i(\frac{pq}{2}+(\vec{n}')_1)T'}\exp\bigr(\frac{\pi i}{2}\vec{n}'^T\mathrm{K}[\vec{k}]\vec{n}'\bigr)\;.
    \end{split}
\end{align}
The sign factor $\e$ in $\bar{S}$ defined as \eqref{eq: Sign factor for WRT S} is considered by $\g\rightarrow \g'$. We should determine $T'(t',i')$ and $\vec{n}'(t',i')$, the arrangement of Bethe vacua in the fixed $a'$ sector of Bethe vacuum-Simple object map \eqref{eq: odd p BVSO map} to finish the modular matrix construction. Since $\bar{S}$, $\bar{T}$ are good modular matrices and $S^{D(\vec{k})},T^{D(\vec{k})}$ will also be good modular matrices due to its TQFT origin, we expect it will be always possible to find $T'$ and $\vec{n}'$ with phase factor $\xi'$ and $\g'\in\{\pm 1\}$ which make $S^{\TFTk}$, $T^{\TFTk}$ not only satisfy the ${\rm SL}(2,\mathbb{Z})$ relation
\begin{align}
    S^2=(ST)^3=C\;,\quad C^2=1\;,\quad C\equiv N^1
\end{align}
but also satisfy other physical conditions introduced after \eqref{eq: S matrix condition starting point}. Note that $\TFTk$ is expected to be a unitary TQFT\cite{Gang:2024wxz}. See appendix~\ref{Ap: Explicit TFTk} for explicit examples.

$t_0$ and $\vec{n}_0$, which correspond to the true Bethe vacuum associated with the trivial object is uniquely determined from positivity condition for the first row. To see this, consider the first column of the $S^{\TFTk}$:
\begin{align} \label{eq: oddp S col1}
    \frac{S^{\TFTk}_{(t',i')(0,1)}}{\gamma'\times2^{-\frac{n}{2}}}=(-1)^{\sum_{k=1}^{n-1}(\vec{n}_{i'})_k+\vec{n}_{i'}\cdot\vec{n}_0+t_0\vec{q}\cdot\vec{n}_{i'}+(1+t_0)t'}.
\end{align}
We want to find $(t_0,\vec{n}_0)$ which makes \eqref{eq: oddp S col1} same for every $(t',i')$. In order to prevent the sign flip in different $t'$, $t_0$ should be $1$. Then \eqref{eq: oddp S col1} becomes
\begin{align} \label{eq: odd p S after choose t0}
    (-1)^{\sum_{k=1}^{n-1}(\vec{n}_{i'})_k+\vec{n}_{i'}\cdot(\vec{n}_0+\vec{q})}
\end{align}
and we can find a unique $\vec{n}_0$ which makes \eqref{eq: odd p S after choose t0} same for every $i'$:
\begin{align}
    (\vec{n}_0)_{i=1,2,\cdots,n-1}=\frac{1}{2}\bigr(1+(-1)^{q_i}\bigr)\;,\quad \vec{n}_0\in(\mathbb{Z}^{n-1}_2,0)\;.
\end{align}
Consequently, $\g'$ is determined as $1$ and the matrix element of $S^{\TFTk}$ is simplified to
\begin{align}
    S^{\TFTk}_{(t',i')(t,i)}=2^{-\frac{n}{2}}(-1)^{\vec{n}_i\cdot\vec{n}'+tT'+\sum_{k=1}^{n-1}(\vec{n}_i)_k+T'\vec{q}\cdot\vec{n}_i+t}\;.
\end{align}
\paragraph{Comment about other candidates for simple objects} We have other candidates $\bigr\{\CO^{(\vec{L},\vec{Q})}\bigr\}$ for the maximally independent set of simple objects. Following the same procedure, we get the $S$ matrix element
\begin{align} \begin{split}
    S^{D(\vec{k})}_{A(a',t',i')A(a,t,i)}=\gamma'&\times\bar{S}_{a'a}
    \\&\times 2^{-\frac{n}{2}}e^{\frac{\pi i}{p}\sum_{k=1}^{n-1}[L_kQ_kq_k((-1)^{t_0}m\vec{n}_{i'}+2(-1)^{T'}\bar{a}(a')\vec{n}_i)_k]}
    \\&\times (-1)^{\vec{n}_i\cdot\vec{n}'+t_0t'+tT'+\sum_{k=1}^{n-1}[(1-L_k)(\vec{n}_i+\vec{n}_{i'})_k]+\vec{n}_{i'}\cdot\vec{n}_0+\vec{q}\cdot(t_0\vec{n}_{i'}+T'\vec{n}_i)+t+t'}\;.
\end{split} \end{align}
To find a constraint for the choice of charge $\vec{Q}$, let's consider the different Weyl quotient for Bethe vacua
\begin{align} \label{eq: Weyl transformation}
    \frac{{\rm BV}^{D(\vec{k})}}{(\textbf{Weyl})}={\rm BV}^{D(\vec{k})}\biggr|_{\vec{G}\neq \vec{0}}\;.
\end{align}
\textit{Assuming} that the Bethe vacuum-Simple object map remains unchanged, we find the Weyl invariance of $S^{D(\vec{k})}$ restricts $\frac{q_w Q_w}{p}\in\mathbb{Z}$ for $w=1,2,\cdots,n-1$ and the resulting $S^{D(\vec{k})}$ becomes the tensor product of the two matrices
\begin{align}
    S^{D(\vec{k})}=\bar{S}\times S^{{\TFTk};\vec{L}}\;,\quad S^{\TFTk;\vec{L}}_{(t',i')(t,i)}=2^{-\frac{n}{2}}(-1)^{\vec{n}_i\cdot\vec{n}'+tT'+\sum_{k=1}^{n-1}[(1-L_k)(\vec{n}_i)_k]+T'\vec{q}\cdot\vec{n}_i+t}
\end{align}
and for each choice of $\vec{L}\in\mathbb{Z}_2^{n-1}$, $t_0$ and $\vec{n}_0$ are uniquely determined as
\begin{align}
    t_0=1\;,\quad (\vec{n}_0)_{i=1,2,\cdots,n-1}=\frac{1}{2}\bigr(1+(-1)^{q_i+L_i}\bigr)\;,\quad \vec{n}_0\in(\mathbb{Z}^{n-1}_2,0)
\end{align}
from the positivity condition for the first row.

\subsubsection{When $p$ is even}
The result of 3-sphere partition function analysis implies that we can write the true Bethe vacuum of the $D(\vec{k})|_A$ theory $V_1$ as
\begin{align} \label{eq: even p possible V1}
    V_1=V\bigr(|p|t_0+(-1)^{t_0}m,\vec{n}_0\bigr)\in\frac{{\rm BV}^{D(\vec{k})}}{(\textbf{Weyl})}\text{ where }t_0\in\mathbb{Z}_2,\;\vec{n}_0\in(\mathbb{Z}_2^{n-1},0)\;.
\end{align}
Here, a unique odd $m\in\{1,2,\cdots,|p|-1\}$ is chosen to satisfy $|q|m\in 2|p|\mathbb{N}_0\pm 1$. Then we can determine the first row of the modular $S$ matrix up to the overall phase factor $\g$ using simple objects \eqref{eq: even p Simples}:
\begin{align}
    S^{D(\vec{k})}_{1A}=\bar{\CO}^{\vec{Q}}_{\CC_+}[A]\bigr|_{V_1}S_{11}^{D(\vec{k})}\;,\quad S^{D(\vec{k})}_{11}=\g\times 2^{-\frac{n-1}{2}}\sqrt{\frac{2}{|p|}}\sin\bigr(\frac{\pi}{|p|}m\bigr)
\end{align}
where $A\in\{1,2,\cdots,(|p|-1)\times2^{n-1}\}$ labels simple objects $\bar{\mathcal{O}}^{\vec{Q}}[\alpha,\vec{n}]$ as
\begin{align}
    A=A(\a,i)\equiv \a+|p|(i-1)\;.
\end{align}
Here, $\a\in\{1,2,\cdots,|p|-1\}$ and $\{\vec{n}_{i=1}\equiv \vec{0},\vec{n}_{i=2,\cdots,2^{n-1}}\}=(\mathbb{Z}_2^{n-1},0)$. In this stage, we can check
\begin{align}
    2^{\frac{n-1}{2}}\times|S^{D(\vec{k})}_{1A(\a=1)}|=|S_{11}|=\sqrt{\frac{2}{|p|}}\sin\bigr(\frac{m\pi}{|p|}\bigr)\;,\quad \frac{S^{D(\vec{k})}_{1A}}{S^{D(\vec{k})}_{1A(\a=1)}}=\frac{S_{1\a}}{S_{11}}
\end{align}
after some computations using \eqref{eq: even p SO values}. $S$ is a modular $S$ matrix of the WRT TQFT determined from the $\HF$ data analysis.\footnote{Although two possible identifications are given by \eqref{eq: even p WRT identification from HF data ver 1} and \eqref{eq: even p WRT identification from HF data ver 2}, the modular $S$ matrix is uniquely determined.} So it is natural to identify
\begin{align}
    \a\longrightarrow\text{WRT index}\;,\quad i\longrightarrow\TFTk\text{ index}\;.
\end{align}
Let's try a Bethe vacuum-Simple object map of form
\begin{align}
    V_{A(\a',i')}\equiv V\bigr(\bar{\a}(\a')\text{ or }\bar{\bar{\a}}(\a'),\vec{n}'\bigr)\in\frac{{\rm BV}^{D(\vec{k})}}{(\textbf{Weyl})}\longleftrightarrow\bar{\CO}^{\vec{Q}}[\a',\vec{n}_{i'}]\;.
\end{align}
Here, $\vec{n}'\in(\mathbb{Z}_2^{n-1},0)$ is function of $i'$ which makes $\{\vec{n}'\}$ is one-to-one with $(\mathbb{Z}_2^{n-1},0)$ under a constraint $\vec{n}'(1)=\vec{n}_0$. $\bar{\a}(\a')$ and $\bar{\bar{\a}}(\a')\equiv |p|-\bar{\a}(\a')$ are rearrangements of $\a'$ defined as \eqref{eq: bar alpha}. Note that $\bar{\a}(1)=m$. So
\begin{align} \label{eq: even p V1 consistency}
    V_1=V(\bar{\a}(1)\text{ or }\bar{\bar{\a}}(1),\vec{n}_0)=V(m\text{ or }|p|-m,\vec{n}_0)
\end{align}
and \eqref{eq: even p V1 consistency} is in agreement with \eqref{eq: even p possible V1}. Then the full modular matrices of the $D(\vec{k})|_A$ are determined by relations \eqref{eq: S matrix dict} and \eqref{eq: T matrix dict}. However, assuming that the Bethe vacuum-Simple object map remains unchanged, we can check that there is no such $Q_1\in2\mathbb{Z}-1$ which makes $S^{D(\vec{k})}$ invariant under the Weyl transformation \eqref{eq: Weyl transformation}. It implies \eqref{eq: even p Simples} cannot be a candidate for a full set of simple objects of the $D(\vec{k})|_A$ theory.

But we still have $|p|-1$ Wilson lines charged under ${\rm SU}(2)^n$
\begin{align} \label{eq: SU(2)n lines for even p}
    \bar{\CO}^{\vec{Q}}[\a=1,2,\cdots,|p|-1,\vec{n}=\vec{0}]=W^{(n,\a-1)}
\end{align}
which pass the criteria \eqref{eq: Simple crit TTI} and using \eqref{eq: SU(2)n lines for even p}, we can construct  modular matrices of $\CD(p,q)|_A$ theory. Note that $\vec{n}=\vec{0}$ lines cannot be simple objects of the $\TFTk$ since their absolute values are not equal to $1$.\footnote{$\alpha=|p|-1$ $\rm{SU}(2)$ line has a unity absolute value but it is more natural to regard it as the simple object of the $\mathcal{D}(p,q)|_A$ theory.} So resulting submatrices $S^{D(\vec{k})}_{\a'\a}$ and $T^{D(\vec{k})}_{\a'\a}$ can be interpreted as modular matrices of the $\CD(p,q)|_A$ theory, multiplied by the vacuum sector element of  modular matrices of the $\TFTk$
\begin{align}
    S^{\TFTk}_{11}=2^{-\frac{n-1}{2}}\overset{\text{Unitary}}{>}0\text{ and }T^{\TFTk}_{11}\;.
\end{align}
By requiring modular matrices to be identified with modular matrices of the WRT TQFT, we find a unique Bethe vacuum-Simple object map
\begin{align}
    V_{\a'}=V(\bar{\a}(\a'),\vec{n}_0)\in\frac{{\rm BV}^{D(\vec{k})}}{(\textbf{Weyl})}\longleftrightarrow W^{(n,\a'-1)}\;.
\end{align}
It implies $V_1=V(m,\vec{n}_0)$ and we find
\begin{align}
    S^{D(\vec{k})}_{\a'\a}=S_{\a'\a}\times2^{-\frac{n-1}{2}}\;,\quad T^{D(\vec{k})}_{\a'\a}=T_{\a'\a}\times T_{11}^{\TFTk}
\end{align}
where $S$ and $T$ are modular matrices of the WRT TQFT corresponding to the identification \eqref{eq: even p WRT identification from HF data ver 1} done by a rearrangement $\bar{\a}$. The other possibility \eqref{eq: even p WRT identification from HF data ver 2} is ruled out.

\subsection{Explicit modular matrices of $\TFTk$ when $p$ is odd}
\label{Ap: Explicit TFTk}
In previous appendix~\ref{Ap: Detail of S matrix computation}, we derived the matrix element of $S^{\TFTk;\vec{L}}$
\begin{align}
    S^{\TFTk;\vec{L}}_{(t',i')(t,i)}=2^{-\frac{n}{2}}(-1)^{\vec{n}_i\cdot\vec{n}'+tT'+\sum_{k=1}^{n-1}[(1-L_k)(\vec{n}_i)_k]+T'\vec{q}\cdot\vec{n}_i+t}
\end{align}
where $t,t'\in\mathbb{Z}_2$ and $i,i'\in\{1,2,\cdots,2^{n-1}\}$. $T'\in\mathbb{Z}_2$ and $\vec{n}'\in(\mathbb{Z}^{n-1}_2,0)$ are functions of $(t',i')$ which makes $\{(T',\vec{n}')\}$ one-to-one with $\bigr(\mathbb{Z}_2,(\mathbb{Z}^{n-1}_2,0)\bigr)$ under constraints
\begin{align}
    T'(0,1)=1\;,\quad \bigr(\vec{n}'(0,1)\bigr)_{k=1,2,\cdots,n-1}=\frac{1}{2}\bigr(1+(-1)^{q_k+L_k}\bigr)\;.
\end{align}
The matrix element of $T^{\TFTk;\vec{L}}$ is
\begin{align}
    T^{\TFTk;\vec{L}}_{(t',i')(t,i)}=\xi'\times \d_{t't}\d_{i'i}e^{\pi i(\frac{pq}{2}+(\vec{n}')_1)T'}\exp\bigr(\frac{\pi i}{2}\vec{n}'^T{\rm K}[\vec{k}]\vec{n}'\bigr)\;,\quad |\xi'|=1\;.
\end{align}
\begin{table}[h]
\centering
\begin{tabular}{c|c|c|c|c}
$n$ & $\vec{k}$   & $(p,q)$     & Sequence of $(T',\vec{n}')$ when $\vec{L}=\vec{0}$& $\xi'$\\ \hline
$3$ & $(2,7,5)$   & $(63,34)$   & $(\langle7\rangle,\langle5\rangle,\langle3\rangle,\langle1\rangle,\langle2\rangle,\langle4\rangle,\langle6\rangle,\langle8\rangle)$  & $e^{\frac{\pi i}{4}+\frac{2\pi i}{3}\mathbb{Z}}$                        \\
$3$ & $(3,5,8)$   & $(109,39)$  & $(\langle6\rangle,\langle4\rangle,\langle2\rangle,\langle8\rangle,\langle7\rangle,\langle1\rangle,\langle3\rangle,\langle5\rangle)$     & $e^{\frac{\pi i}{12}+\frac{2\pi i}{3}\mathbb{Z}}$                        \\
$4$ & $(4,3,6,2)$           & $(113,31)$        & \makecell{$(\langle10  \rangle,\langle 8 \rangle,\langle 14 \rangle,\langle 4 \rangle,\langle  2\rangle,\langle 16 \rangle,\langle 6 \rangle,\langle 12 \rangle,$\\$\langle13  \rangle,\langle 3 \rangle,\langle 9 \rangle,\langle 7 \rangle,\langle 5\rangle,\langle 11 \rangle,\langle 1 \rangle,\langle 15 \rangle)$}                               & $e^{\frac{2\pi i}{3}\mathbb{Z}}$         
\end{tabular}
\captionof{table}{Examples of determined $T'$ and $\vec{n}'$. In order to express results explicitly, we defined $\vec{n}_{i=1,2,\cdots,2^{n-1}}\equiv((i-1)_2,0)$, $(t,i)\equiv i+t\times 2^{n-1}$ and $(t,\vec{n}_i)\equiv\langle i+t\times 2^{n-1}\rangle$. Note that $\vec{L}=\vec{0}$ is the case when all of simple objects are $\mathrm{SU}(2)$ Wilson lines.}
\label{tab: odd p TFTk BVSO map eg}
\end{table}
We expect that it will always be possible to find unique $T'$ and $\vec{n}'$ which make $S^{\TFTk;\vec{L}}$ and $T^{\TFTk;\vec{L}}$ behave as unitary modular matrices which satisfy conditions introduced after \eqref{eq: S matrix condition starting point}. With determined $T'$ and $\vec{n}'$, the phase factor $\xi'$ is also determined up to $e^{\frac{2\pi i}{3}\mathbb{Z}}$.
For the cases introduced in table~\ref{tab: odd p TFTk BVSO map eg}, we checked it is true. Furthermore, we also observed that $S^{\TFTk;\vec{L}}$ is independent of the choice of $\vec{L}$ for cases in table~\ref{tab: odd p TFTk BVSO map eg}.

\paragraph{Case 1} $\mathrm{TFT}[(2,7,5)]$
\\\\
\noindent $S^{\mathrm{TFT}}=\frac{1}{2\sqrt{2}}\begin{pmatrix}
        1&1&1&1&1&1&1&1\\
        1&1&-1&-1&1&1&-1&-1\\
        1&-1&1&-1&-1&1&-1&1\\
        1&-1&-1&1&-1&1&1&-1\\
        1&1&-1&-1&-1&-1&1&1\\
        1&1&1&1&-1&-1&-1&-1\\
        1&-1&-1&1&1&-1&-1&1\\
        1&-1&1&-1&1&-1&1&-1
    \end{pmatrix}$
\\\\
\noindent $T^{\mathrm{TFT}}=e^{\frac{\pi i}{4}+\frac{2\pi i}{3}\mathbb{Z}}\mathrm{diag}\{-1,-1,-1,1,-i,-i,i,-i\}$

\paragraph{Case 2} $\mathrm{TFT}[(3,5,8)]$
\\\\
\noindent $S^{\mathrm{TFT}}=\frac{1}{2\sqrt{2}}\begin{pmatrix}
        1&1&1&1&1&1&1&1\\
        1&1&1&1&-1&-1&-1&-1\\
        1&1&-1&-1&-1&-1&1&1\\
        1&1&-1&-1&1&1&-1&-1\\
        1&-1&-1&1&1&-1&-1&1\\
        1&-1&-1&1&-1&1&1&-1\\
        1&-1&1&-1&-1&1&-1&1\\
        1&-1&1&-1&1&-1&1&-1
    \end{pmatrix}$
\\\\
\noindent $T^{\mathrm{TFT}}=e^{\frac{\pi i}{12}+\frac{2\pi i}{3}\mathbb{Z}}\mathrm{diag}\{1,-1,i,-i,1,1,-i,-i\}$

\setcounter{MaxMatrixCols}{16}
\paragraph{Case 3} $\mathrm{TFT}[(4,3,6,2)]$
\\\\
\noindent
$S^{\mathrm{TFT}}=\frac{1}{4}\begin{pmatrix} 1 & 1 & 1 & 1 & 1 & 1 & 1 & 1 & 1 & 1 & 1 & 1 & 1 & 1 & 1 & 1 \\
 1 & 1 & 1 & 1 & 1 & 1 & 1 & 1 & -1 & -1 & -1 & -1 & -1 & -1 & -1 & -1 \\
 1 & 1 & 1 & 1 & -1 & -1 & -1 & -1 & 1 & 1 & 1 & 1 & -1 & -1 & -1 & -1 \\
 1 & 1 & 1 & 1 & -1 & -1 & -1 & -1 & -1 & -1 & -1 & -1 & 1 & 1 & 1 & 1 \\
 1 & 1 & -1 & -1 & -1 & -1 & 1 & 1 & -1 & -1 & 1 & 1 & 1 & 1 & -1 & -1 \\
 1 & 1 & -1 & -1 & -1 & -1 & 1 & 1 & 1 & 1 & -1 & -1 & -1 & -1 & 1 & 1 \\
 1 & 1 & -1 & -1 & 1 & 1 & -1 & -1 & -1 & -1 & 1 & 1 & -1 & -1 & 1 & 1 \\
 1 & 1 & -1 & -1 & 1 & 1 & -1 & -1 & 1 & 1 & -1 & -1 & 1 & 1 & -1 & -1 \\
 1 & -1 & 1 & -1 & -1 & 1 & -1 & 1 & 1 & -1 & 1 & -1 & -1 & 1 & -1 & 1 \\
 1 & -1 & 1 & -1 & -1 & 1 & -1 & 1 & -1 & 1 & -1 & 1 & 1 & -1 & 1 & -1 \\
 1 & -1 & 1 & -1 & 1 & -1 & 1 & -1 & 1 & -1 & 1 & -1 & 1 & -1 & 1 & -1 \\
 1 & -1 & 1 & -1 & 1 & -1 & 1 & -1 & -1 & 1 & -1 & 1 & -1 & 1 & -1 & 1 \\
 1 & -1 & -1 & 1 & 1 & -1 & -1 & 1 & -1 & 1 & 1 & -1 & -1 & 1 & 1 & -1 \\
 1 & -1 & -1 & 1 & 1 & -1 & -1 & 1 & 1 & -1 & -1 & 1 & 1 & -1 & -1 & 1 \\
 1 & -1 & -1 & 1 & -1 & 1 & 1 & -1 & -1 & 1 & 1 & -1 & 1 & -1 & -1 & 1 \\
 1 & -1 & -1 & 1 & -1 & 1 & 1 & -1 & 1 & -1 & -1 & 1 & -1 & 1 & 1 & -1 \\
\end{pmatrix}$
\\\\
\noindent $T^{\mathrm{TFT}}=e^{\frac{2\pi i}{3}\mathbb{Z}}\mathrm{diag}\{i,i,-i,-i,-1,-1,-1,-1,i,-i,-i,i,1,-1,1,-1\}$

\section{Quantum dilogarithm function}
\label{Ap: QDL}
On the squashed 3-sphere background, contribution of matter called the ``Tetrahedron theory" $\mathcal{T}_\Delta$\cite{Dimofte:2011ju} defined as the theory of $3$d $\mathcal{N}=2$ free chiral multiplet $\Phi$ with the UV Chern-Simons term of level $-\frac{1}{2}$ associated with the background vector multiplet gauging its global $\mathrm{U}(1)$ symmetry $\Phi\rightarrow e^{i\theta}\Phi$, is given by the quantum dilogarithm function $\psi_{\hbar}(Z)$ defined as
\begin{align}
    \psi_{\hbar=2\pi ib^2} (Z)\equiv \begin{cases}
        \prod_{r=1}^\infty \frac{1-q^r e^{-Z}}{1-\bar{q}^{-r+1}e^{-Z/b^2}} &(|q|<1)
        \\\prod_{r=1}^\infty \frac{1-\bar{q}^r e^{-Z/b^2}}{1-q^{-r+1}e^{-Z}} &(|q|>1)
    \end{cases}
\end{align}
where $q\equiv e^{2\pi ib^2}$ and $\bar{q}\equiv e^{2\pi ib^{-2}}$. Here, $Z$ is the ($2\pi b$-multiplied) real mass parameter of the background vector multiplet. Note that half-integral choice of the Chern-Simons level at UV gives properly quantized Chern-Simons level in IR. See \cite{Closset:2018ghr} for more details.
\paragraph{$b^2\rightarrow 0$ expansion}
With modified polylogarithm functions defined with the floor function and the unit step function $\theta$
\begin{align}
\label{eq: modified Li}
\begin{aligned}
\tilde{\mathrm{Li}}_1(e^{-Z})\equiv& \mathrm{Li}_1(e^{-Z})+2\pi i\biggr\lfloor\frac{\mathrm{Im}(Z)}{2\pi}\biggr\rfloor\times\theta\bigr(-\mathrm{Re}(Z)\bigr)
\\=&-\mathrm{Log}(1-e^{-Z})+2\pi i\biggr\lfloor \frac{\mathrm{Im}(Z)}{2\pi}\biggr\rfloor\times\theta\bigr(-\mathrm{Re}(Z)\bigr)\;,
\\\tilde{\mathrm{Li}}_2(e^{-Z})\equiv& \mathrm{Li}_2(e^{-Z})-2\pi i\biggr\lfloor\frac{\mathrm{Im}(Z)}{2\pi}\biggr\rfloor\biggr(Z-\pi i \biggr(\biggr\lfloor \frac{\mathrm{Im}(Z)}{2\pi}\biggr\rfloor+1\biggr)\biggr)\times\theta\bigr(-\mathrm{Re}(Z)\bigr)
\end{aligned}
\end{align}
which have the singularity structure same with the $\log\psi_\hbar(Z)$\footnote{See figure 9 in~\cite{dimofte:2012abab}.}, asymptotic behavior of $\psi_\hbar$ when $b^2\rightarrow0
$ is known as\cite{Gang:2025,Gang:2021hrd}
\begin{align}
    \log \psi_{\hbar}(Z)\xrightarrow{b^2\rightarrow0}\frac{1}{\hbar}\tilde{\mathrm{Li}}_2(e^{-Z})+\frac{1}{2}\tilde{\mathrm{Li}}_1(e^{-Z})+\mathcal{O}(\hbar)\;.
\end{align}

\bibliographystyle{JHEP}
\bibliography{biblio.bib}
\end{document}